\documentclass[setspace,amsmath,amssymb,fleqn]{article}
\usepackage{setspace}
\usepackage{graphicx}
\usepackage{amsmath}
\usepackage{amssymb}
\usepackage[all]{xy}
\usepackage{rsfs}
 
\newcommand{\bk}{{\boldsymbol{k}}}

\newcommand{\ra}{\rangle}
\newcommand{\la}{\langle}
\begin{document}
\begin{titlepage}

\begin{flushright}
\today
\end{flushright}

\vspace{1in}

\begin{center}

{\bf Two resonant quantum electrodynamics models of quantum measuring systems}

\vspace{1in}

\normalsize

\renewcommand\thefootnote{\fnsymbol{footnote}}

{Eiji Konishi\footnote[0]{E-mail address: konishi.eiji.27c@kyoto-u.jp}
}

\normalsize
\vspace{.5in}

{\it Graduate School of Human and Environmental Studies,\\
 Kyoto University, Kyoto 606-8501, Japan}
\end{center}

\vspace{1in}

\baselineskip=24pt
\begin{abstract}
A quantum measurement scheme is suggested in two resonant models of quantum electrodynamics.
The first model is the brain, where, for the propagation of its action potentials, the free electron laser-like coherence mechanism recently investigated by the author is comprehensively applied.
The second model is an assembly of Preparata et al.'s
coherence domains, in which we incorporate the quantum field theory of memory advocated by Umezawa et al.
These two models are remarkably analogous.
\end{abstract}

\vspace{.1in}

{\it Keywords}: Measurement problem; quantum optics; resonance; coherence.

\vspace{.6in}

\end{titlepage}

\section{Introduction}
The quantum electrodynamics (QED) of matter and radiation has various coherence mechanisms above the structure of its Hamiltonian due to the cooperative quantum state of radiators and collective instability in a many-body system with a long-range wave-particle interaction \cite{Dicke,FEL,Bonifacio,Nature}.

In this paper, we treat two resonant QED models of such a mechanism and discuss the quantum measurement scheme of these models based on the quantum coherence mechanism of each.

The first model is a free electron laser (FEL)-like model for resonant systems of ion-solvated water and radiation photons which was recently investigated by the author \cite{PLA1,PLA2,arXiv1}.

We comprehensively apply this model to the system for the propagation of an action potential mediated by the electric charge currents of water-solvated sodium ions Na$^+$ in the myelinated neuronal axons in the neural network of the brain.

The second model is the Preparata
et al.'s
model of superradiance in a coherence domain \cite{Preparata,Enz}.\footnote{For Ref. \cite{Preparata}, see also Refs. \cite{Preparata2,GPV,Preparata3}.}

Preparata
et al.
showed that a superradiant phase transition occurs with {\it no} cavity and {\it no} pump, if the number density of atoms or molecules resonantly interacting with the radiation in a domain that has the dimension of the resonance wavelength exceeds a threshold and the temperature is below a critical value \cite{Dicke,Preparata,Enz,HL1,HL2,NoGo1,NoGo2}.
Remarkably, the ground state after this phase transition is a non-perturbative one, and its energy is less than that of the perturbative ground state.

We regard this Preparata
et al.'s
coherence domain as an {\it atomic spatial region of coherent and homogeneous evolution in the resonant QED nature}.\footnote{In this paper, {\it resonant QED nature} refers to the {\it absolutely open quantum system based on only resonant QED}.}
Namely, we can construct the complete non-perturbative ground state, starting from a single coherence domain, by nucleating the appropriate number of coherence domains.
In this picture, the resonant QED nature is an aggregation of a myriad of fundamental coherence domains \cite{Preparata}.

We apply this Preparata
et al.'s
model to an assembly of coherence domains by incorporating the quantum field theory of memory advocated by Umezawa et al. \cite{Umezawa,RU,STU1,STU2,Vitiello,JPY}.
The fundamental processes of memory are {\it writing}, {\it retrieval} and {\it reading}.

In this paper, we show how to express states in each model by using classical bits.
However, we do not incorporate information processing, that is, changes of these states in the {\it system of causally interacting discrete elements} (i.e., the {\it discrete dynamical system}) in each model.
Information processing is a higher-order activity than the focus of the present investigation, which we specifically limit to the study of quantum measurement processes (i.e., information transduction) only.

The goal of this paper is conceptual.
It is to find a possible way to embed a quantum measurement scheme, which includes the event reading step, in two discrete dynamical systems, namely, the brain, in the first model, and the resonant QED nature, in the second model, while maintaining the consistency of this scheme with the informational structure of each system.

The plan of this paper is as follows.
In Sects. 2 and 3, we study the quantum measurement scheme of the brain that occurs via the sensory organs and the scheme of the assembly of coherence domains.
In Sect. 4, we explain how these two systems are analogous, and we consider a thought experiment.
In ``Appendix A'', we derive the decoherence mechanism invoked in the first model.
In ``Appendix B'', we give a brief account of an informational interpretation of the second model.

\section{The first model: the brain}

\subsection{FEL-like resonant model}

First, we briefly review the results in Refs. \cite{PLA1,PLA2,arXiv1}.

\subsubsection{The resonant system of photons and ion-solvated water}

We consider a model for a resonant system of radiation photons and ion cluster-solvated ordered rotating water molecules, in which ions in the cluster carry the same electric charge and move with very low, non-relativistic velocities $v\ll c$ in a direction parallel to a static electric vector field $E_0$ applied in a single $z$-direction.
Here, the bulk water molecules that screen the charges of ions do not form part of the following laser mechanism; the bulk water molecules are in a {\it mixed state}, assumed to be the canonical distribution.

In this model, the dimensions of the system of one ion and its solvent water molecules fall within the resonant wavelength of radiation.
So, the time evolution of this system is symmetrized with respect to permutations of water molecules related to this ion.
It is further assumed that the dimensions of the ion cluster are much shorter than this wavelength.

In a seminal paper, Ref. \cite{GPV}, it was shown that, due to the resonant interactions between the water molecules and the radiation field, the static electric vector field $E_0$ that couples with the electric dipole moments of water molecules and mixes their rotational states induces, in the limit cycle of the system, a {\it permanent} electric polarization of water molecules in the $z$-direction \cite{GPV}.

Using this result, in Refs. \cite{PLA1,arXiv1} we combined Dicke {\it superradiation} with the lowest cooperation number (here, obtained by exciting and de-exciting the ground state and the rotationally symmetric first excited state of water molecules, respectively, in the limit cycle of the system starting from the thermal equilibrium state) \cite{Dicke,GPV} for the time-dependent closed system of the $30-35$ water molecules solvating each ion \cite{SB} in a {\it pure state} with a wave-particle interaction between the transverse electro-magnetic field radiated from the rotating water molecules and the ion-solvated water molecules.
We obtained this wave-particle interaction as the minimal coupling of the transverse electro-magnetic field and the electric dipole current of the ion-solvated water molecules.

In our resonant system, to a good approximation, the radiation field exchanges energy with water molecules only through excitation and de-excitation between the two lowest levels of the internal rotation of the hydrogen atoms of a water molecule, which is considered as a {\it quantum mechanical rigid rotator} with a subtly {\it variable} moment of inertia around its electric dipole axis.
The energy difference between these two levels for the {\it average} moment of inertia is ${\varepsilon}_{\rm w}$, where ${\varepsilon}_{\rm w}/(\hbar c) \approx 160$ [cm$^{-1}$].

In this two-level approximation of the rotational spectrum of water molecules, the Hilbert space of the internal rotational states of each water molecule is four-dimensional and is spanned by the state $|l,m\ra$ with $l=0$ and $m=0$ together with the three states $|l,m\ra$ with $l=1$ and $m=1,0,-1$.

In the framework of Refs. \cite{PLA1,PLA2,arXiv1}, we reduce this Hilbert space to a two-dimensional space by characterizing water molecules by referring to their energy spin operators\footnote{In this paper, ${\rm i}$ in the roman typeface denotes $\sqrt{-1}$, and hatted variables are quantum operators.}
\begin{eqnarray}
\widehat{s}^1&=&\frac{1}{2}[|e\ra\la g|+|g\ra\la e|]\;,\label{eq:Spin1}\\
\widehat{s}^2&=&\frac{1}{2{\rm i}}[|e\ra\la g|-|g\ra\la e|]\;,\label{eq:Spin2}\\
\widehat{s}^3&=&\frac{1}{2}[|e\ra\la e|-|g\ra\la g|]\;.\label{eq:Spin3}
\end{eqnarray}
The energy spinor in the two-dimensional energy state space is spanned by the {\it ground} state $|g\ra\equiv|0,0\ra$ and the {\it excited} energy state $|e\ra\equiv|1,1\ra$.
These energy spin operators obey an $su(2)$ algebra $[\widehat{s}^i,\widehat{s}^j]={\rm i}\epsilon_{ijk}\widehat{s}^k$.

For each water molecule, we introduce a Cartesian frame with basis $(e_1,e_2,e_3)$.
Its electric dipole moment direction vector is taken to be $e_3$, and we choose the quantization axis of the angular momentum of that water molecule so that it lies along $e_3$.
We can choose the angle of rotation of $(e_1,e_2)$ arbitrarily on the plane to which $e_3$ is normal by adjusting the arbitrary phase associated with the energy spin eigenstates $|g\ra$ and $|e\ra$ \cite{AE}.

For a single rotating water molecule, the electric dipole moment operator vector ${\widehat{d}}$ is a spatial vector operator, and thus has odd-parity and no non-zero diagonal matrix elements.
In the original four-dimensional Hilbert space, this electric dipole moment operator vector ${\widehat{d}}$ does not interchange any two different states $|1,m_1\ra$ and $|1,m_2\ra$, and its three component operators $\widehat{d}^1$, $\widehat{d}^2$ and $\widehat{d}^3$ interchange the ground state $|0,0\ra$ with the states $-(|1,1\ra-|1,-1\ra)/\sqrt{2}$, $i(|1,1\ra+|1,-1\ra)/\sqrt{2}$ and $|1,0\ra$, respectively.

In the truncated two-dimensional Hilbert space, the truncated electric dipole moment operator vector ${\widehat{d}}_{tr}$ can be written as
\begin{equation}
{\widehat{d}}_{tr}=e_1(-\widetilde{d}_0\widehat{s}^1)+e_2(-\widetilde{d}_0\widehat{s}^2)+e_3\widehat{0}
\end{equation}
with a constant $\widetilde{d}_0$.

Now, we construct the Hamiltonian of each water molecule in the radiation gauge.

First, for each water molecule with the average moment of inertia, the free Hamiltonian is
\begin{equation}
\widehat{H}_{\rm free}=\frac{1}{2}{\varepsilon}_{\rm w}\widehat{I}_{3,1}\;,
\end{equation}
where
\begin{equation}
\widehat{I}_{3,1}\equiv [|1,1\ra\la 1,1|+|1,0\ra\la 1,0|+|1,-1\ra\la 1,-1|-|0,0\ra\la 0,0|]\;.
\end{equation}

Next, the semi-classical Hamiltonian for the interaction between the classical radiation ${A}$, of the specific mode, and each water molecule is
\begin{equation}
\widehat{H}_{\rm int}=-{A}\cdot \dot{{\widehat{d}}}_{tr}\;.
\end{equation}

Whereas the original electric dipole moment operator vector ${\widehat{d}}$ transforms as a {\it vector} with respect to spatial rotations, the operator $\widehat{I}_{3,1}$ is a {\it scalar} operator; since $\widehat{I}_{3,1}$ is a diagonal operator, it cannot be a component of a spatial vector.
So, we truncate the electric dipole moment operator in the interaction Hamiltonian but do not truncate the free Hamiltonian.
Note that $[\widehat{I}_{3,1}/2,\widehat{s}^a]=[\widehat{s}^3,\widehat{s}^a]$ holds for $a=1,2,3$.

\subsubsection{Dynamical mechanism for coherence}

In this resonant system, coherent and collective behavior of the $XY$-phases of ion-solvated water molecules over all ions is generated.

Here, the {\it ion-solvated water molecule's $XY$-phase} is defined such that the angular frequency in this phase is proportional to the variation of the moment of inertia, where the proportionality constant is determined by the resonance condition.
This phase is the $XY$-phase of the energy spin minus $\omega_ct$ for the resonance angular frequency $\omega_c$.

The mechanism for this behavior of the $XY$-phases consists of two interlocked parts in a positive feedback cycle \cite{PLA1}.

\begin{enumerate}

\item[Part 1] The first part is the ordering of the electric dipole moment direction vectors of water molecules, permanently polarized in the $z$-direction, that solvate each ion moving along the $z$-axis.
In this, the static electric field is applied towards the positive $z$-direction.

As a consequence, the radiation from the clusters of ordered rotating water molecules is almost monochromatic, and the time-dependent process of the radiation field and order of water molecules approximates a coherent wave amplified along the $z$-axis.

\item[Part 2]

The approximations in the first part are improved by positive feedback in the second part of the mechanism.

In this second part, exponential instability of the fluctuation around the dynamic equilibrium state, which is our ready state, accompanies both the magnification of the radiation intensity and the ion-solvated water molecule's $XY$-phase bunching \cite{Bonifacio,Kim}.

Indeed, in this second part of the mechanism, the equations of motion of the $XY$-phase of the ion-solvated water in the superradiant pure state, with respect to each solvated ion, and the transverse electromagnetic field of the system can be expressed in terms of a conventional FEL system.
(The $XY$-phase of the energy spin of the bulk water in the mixed state is uniformly random.)

\end{enumerate}

As a result, this positive feedback cycle leads to dynamical coherence over all ions and the radiation field, induced by collective instability in the wave-particle interaction, and the bunching process of the system will saturate according to the results of a steady state FEL model \cite{Nature}.

\subsection{Action potentials}
In the rest of this section, we consider the resonant QED system in the human brain.

\subsubsection{Description of action potentials}

The fundamental ingredients of the neural network in the human brain are {\it neurons} (i.e., neural cells) of which there are $\sim 10^{11}$, and the associated {\it synapses}.
A synapse is a junctional structure between two neurons, and each neuron has about $10^3$--$10^4$ synapses.
The activity of the synapses is controlled by electric or chemical signals \cite{Kandel}.

The classical physical definite formulation of the activity of neurons in the brain is based on the Hodgkin-Huxley model \cite{Kandel,HH,Hodgkin}.

In this model, the cell membrane and the ion channels of a neuron are regarded as the condenser and dynamical registers, respectively, in an electric circuit along the axial direction inside and outside of the neuron separated by the cell membrane.

In neuronal cell membranes, voltage-dependent sodium (Na$^+$) and potassium (K$^+$) ion channels are embedded.
Together with these and other ion channels and ion transporters, the neuronal cell membrane (the {\it axon} membrane) maintains an electric potential difference $U_0 \approx -0.07$ [V] across itself by adjusting the concentrations of ions (mainly, K$^+$, Na$^+$ and Cl$^-$) inside and outside of the neuron.
This electric potential difference arises from the equilibrium between the K$^+$-concentration gradient diffusion force (from the inside to the outside of the neuron) and the electric gradient Coulomb force on K$^+$ (from the outside to the inside of the neuron) as a consequence of the diffusion of positive electric charges, K$^+$, from inside to outside of the neuron.
At the same time, there is more K$^+$ inside of the neuron than outside, and so potassium ions diffuse from inside the neuron to outside through the K$^+$-selective pores until equilibrium is reached, and sodium ions are transported out of the neuron by the sodium-potassium exchange pump.
We call this neural state the {\it resting state}.

When the membrane electric potential exceeds a negative threshold value, the voltage-dependent sodium ion-channels in the axon membrane open up, which induces sodium ions to flow into the axon.
This rapid depolarization of the membrane electric potential, called an {\it action potential}, propagates down the axon, as a chain reaction, changing the membrane electric potential difference to a value $U_1\approx 0.03$ [V] until it reaches the terminals of the neuron, that is, the pre-synaptic sites.
We call this neural state the {\it firing state}.

After generating an action potential, the membrane electric potential repolarizes and returns to the resting state by the inactivation of the sodium ion-channel and the activation of the potassium ion-channel.
This occurs within a few milliseconds.

The action potential, as an electrical signal, is then converted to pulse form membrane electric potential inputs to the other neurons through neurotransmitters, that is, as an excitatory (if inputs are positive) or inhibitory (if inputs are negative) synaptic transmission from the original neuron to the other neurons.

Due to the threshold structure (i.e., that the rule is {\it all-or-nothing}) for the accumulated membrane electric potential inputs (i.e., the accumulated changes in ion concentrations) to generate a new action potential, the neural network can be characterized as a non-linear many-body system.

\subsubsection{Application of the FEL-like mechanism}

Now, we apply our FEL-like mechanism to the system for the propagation of an action potential mediated by the electric charge currents of water-solvated sodium ions Na$^+$ in typical myelinated (i.e., insulated by myelin sheathing) neuronal axons of the human brain \cite{arXiv1,Kandel,Hodgkin,Lodish}.

The typical diameter $l_a$ of central nervous system axons is $10$ [$\mu$m] \cite{Axon}, which is much shorter than the resonant wavelength of radiation $l_c\equiv hc/{\varepsilon}_{\rm w}\approx 400$ $[\mu{\rm m}]$.
This characteristic length $l_c$ is the so-called {\it coherence length} (i.e., the wavelength of a resonant photon).
The total number of sodium ions that migrate in during an action potential event is estimated to be $Nn_{{\rm ms}}\sim 10^6$ \cite{Tegmark}, where $n_{{\rm ms}}$ (estimated to be between $50$ and $100$ \cite{Kandel}) is the number of myelin sheaths on one myelinated axon with an assumed length of $10$ [cm].
The conduction velocity $v$ of action potential propagation along a myelinated axon is up to $150$ [m$\cdot$s$^{-1}$] \cite{Hodgkin}.
When we approximate $E_{0,z}$ to be uniform in the $z$-direction along each myelin sheath, with run length $l_r\sim 1$ [mm] \cite{Lodish} for the electric potential sloping toward the $z$-direction, it is estimated to be $\Delta U/1$ [mm] $\approx 100$ [V$\cdot$m$^{-1}$] for an electric potential difference of $\Delta U\equiv U_1-U_0\approx 0.1$ [V] between the neural firing and resting states \cite{Axon}.

The concluding formulae for the steady-state regime of the FEL-like mechanism are \cite{arXiv1}
\begin{eqnarray}
A_0&=&c_A\cdot \rho^{2/3}\cdot P^{1/3}_z\;,\label{eq:FA}\\
c_A&\approx&2.6\cdot 10^{-22}\ [{\rm m}^3\cdot {\rm kg}\cdot{\rm s}^{-2}\cdot {\rm A}^{-1}]\;,\\
t&=&c_t\cdot \rho^{-1/3}\cdot P^{-2/3}_z\;,\label{eq:FT}\\
c_t&\approx&8.1\cdot 10^{-5}\ [{\rm m}^{-1}\cdot{\rm s}]\;.
\end{eqnarray}
Here, $\rho=N/V$ (where $V$ is the volume of the system) is the sodium ion number concentration in the system and $P_z$ is the permanent electric polarization of water molecules under the static electric field $E_{0,z}$.
The first formula gives the saturated value of the transverse electro-magnetic field modulus $A_0$ in the radiation gauge, and the next formula gives the gain time.

In this steady-state regime, coherent dynamics of the radiation field and maximally bunched phases of water molecules arises: the quantum coherence of water molecules is coupled over the system of sodium ion-solvated water; the radiation field is coherent and its intensity is magnified by a multiplicative factor of the order of $N^{4/3}$ \cite{PLA1,arXiv1}.

In this system, by setting $V \approx \pi l_a^2l_r/4$ and $P_z\approx 4.9\cdot 10^{-7}$ according to the formulae in Ref. \cite{GPV}, we obtain \cite{arXiv1}
\begin{eqnarray}
A_0&\approx&5.1\cdot 10^{-13}\ [{\rm m}\cdot {\rm kg}\cdot {\rm s}^{-2}\cdot{\rm A}^{-1}]\;,\label{eq:NUM1}\\
t&\approx& 2.6\cdot 10^{-6}\ [{\rm s}]\;.\label{eq:NUM2}
\end{eqnarray}
Here, the gain time is of the order of the dynamical time scale of action potential propagation $1\ [{\rm mm}]/v\approx 6.7\cdot 10^{-6}$ [s], so the quantum coherence effect is relevant to this system.

\subsection{Sensory organs and sensory transduction}

Next, using the result from Sect. 2.2, we model the quantum measurement process occurring in the brain via the sensory organs ({\it ear}, {\it eye}, {\it skin}, etc.).

\subsubsection{Overview of the measurement process}

It is broadly accepted as fact that external stimuli received by the human brain are coded via information transduction (the {\it coding process}) in which {\it the resting/firing state is expressed by one classical bit for each neuron} \cite{Kandel}.
As quantitatively described in Sect. 2.2, an action potential is, in total, an event that occurs at a classical mechanical scale.
Thus, one cannot suppose that the brain carries out a totally {\it quantum} coding process in which the superpositions of the resting/firing state of each neuron are expressed by a qubit: the human brain is {\it not} a quantum computer.

This no-go statement is reinforced by the short decoherence time $t_{{\rm dec}}\sim 10^{-20}$ [s] of the spatial superposition of sodium ions in the neural firing state $|{\cal F}\ra$ and resting state $|{\cal R}\ra$, calculated in Ref. \cite{Tegmark}, such that
\begin{eqnarray}
(\alpha_1|{\cal F}\ra+\alpha_2|{\cal R}\ra)(\bar{\alpha}_1\la {\cal F}|+\bar{\alpha}_2\la {\cal R}|)
\overset{t_{\rm dec}}{\longrightarrow}|\alpha_1|^2|{\cal F}\ra\la{\cal F}|+|\alpha_2|^2|{\cal R}\ra\la {\cal R}|\;.\label{eq:dec}
\end{eqnarray}

For this reason, in the {\it measurement process} occurring in the brain via the sensory organs, the decohered quantum entanglement with an external stimulus ${\cal O}$ is to be kept by the quantum states of the corresponding codes $\{{\cal F}\}$ themselves until the measurement is completed and the state reduction for the external stimulus ${\cal O}$ ({\it sound}, {\it light}, {\it force}, {\it heat}, etc.) occurs.
(Here, the external stimulus is translated into neural firing states of neurons via processing by the sensory organs.)
Namely, in our modeling, the measurement process is {\it partially quantum}: the coding process is {\it classical}, and the propagation of signals (that is, action potentials) relies on {\it quantum} coherence.

\subsubsection{Mathematical model: states and variables}

Now, we start the specific description.

We use $|{\cal O}_1\ra$, $|{\cal O}_2\ra$,$\ldots$ to denote the non-degenerate eigenstates of a discrete observable $\widehat{{\cal O}}$ in the state space ${\mathfrak H}_{st}$ of the system, $S_0$, of the external stimulus.

The state space ${\mathfrak H}_{br}$ of the brain is the tensor product of the state space ${\mathfrak H}_{so}$ of the macroscopic sensory organ in question and the state spaces $\bigl\{{\mathfrak H}^{(\nu)}_n\bigr\}$ of neurons $\{\nu\}$:
\begin{eqnarray}
{\mathfrak H}_{br}&=&{\mathfrak H}_{so}\otimes \Biggl(\bigotimes_{\nu}{\mathfrak H}^{(\nu)}_n\Biggr)\;,\\
{\mathfrak H}^{(\nu)}_n&\subset&{\mathfrak{H}}^{(\nu)}_{{\rm Na}^+}\otimes{\mathfrak{H}}^{(\nu)}_{{\rm H_2O}}\;.
\end{eqnarray}

Here, the state spaces $\bigl\{{\mathfrak H}^{(\nu)}_n\bigr\}$ depend on the neural firing state $|{\cal F}\ra$ and the resting state $|{\cal R}\ra$.

Specifically, the pure state part of the neural firing state of a neuron takes the form
\begin{equation}
|{\cal F}\ra=\bigotimes_{I=1}^N|{\rm WP}_I^{{\rm Na}^+};v\ra|{\rm SR}_I^{\rm H_2O};\theta_I\ra\;.\label{eq:Firing}
\end{equation}
Here, $|{\rm WP}_I^{{\rm Na}^+};v\ra$ is a wave packet state of the $I$-th sodium ion ($I=1,2,\ldots,N$) migrated by the action potential with conduction velocity $v$, and $|{\rm SR}_I^{\rm H_2O};\theta_I\ra$ is a superradiant state with the lowest cooperation number and phase $\theta_I$ for the water molecules that solvate this sodium ion.

As explained in Sect. 2.1, after the gain time elapses, coherence among all phases $\theta_I$ ($I=1,2,\ldots,N$) in Eq. (\ref{eq:Firing}) is realized within the coherence domain of the radiated photons.

Next, ideally, the pure state part of the resting state $|{\cal R}\ra$ of a neuron takes the form the sodium ion empty state:
\begin{equation}
|{\cal R}\ra=|\emptyset^{{\rm Na}^+}\ra\;.
\end{equation}

We use $|{\mathfrak A}_0\ra$ to denote the quantum state of the macroscopic sensory organ, $A_{so}$, in ${\mathfrak H}_{so}$ that couples to the state $|{\cal O}\ra$ in ${\mathfrak H}_{st}$ to form a direct-integral mixture (see Eq. (\ref{eq:element})) of a respective element of {\it continuous superselection sectors} (that is, simultaneous eigenspaces of the continuous superselection rule observables in the combined system $S_{0}+A_{so}$), ${\mathfrak H}_{st}(\overline{P_I})\equiv{\mathfrak H}_{st}$, in
\begin{equation}
{\mathfrak H}_{st}\otimes {\mathfrak H}_{so}=\int^\bigoplus {\mathfrak H}_{st}(\overline{P_I})\prod_Id\overline{P_I}\;,\label{eq:Hst}
\end{equation}
where $I$ labels the sensory cells.

Here, $\overline{\widehat{P}_I}$ is the {\it continuous superselection rule observable} \cite{Araki1,Araki2} of the $I$-th sensory cell which is characterized by the properties of having a virtually continuous spectrum (i.e., being able to be very sharply measured) and commuting with (i.e., being able to be simultaneously measured with) all observables of the sensory cell\footnote{Because of the latter property, for two state vectors belonging to different continuous superselection sectors, the corresponding matrix elements of {\it any} observable of the combined system $S=S_0+A_{so}$ are always zero.
From this fact, the direct-integral structure of the density matrix of $S$ (see Eqs. (\ref{eq:element0}) and (\ref{eq:element1})), and the continuous superselection rule of the observables of $S$ in Eq. (\ref{eq:A8}) follow.}, where the sensory cell is treated as a quantum mechanical object.

Specifically, each of continuous superselection rule observables $\Bigl\{\overline{\widehat{P}_j^{(i)}}\Bigr\}\equiv\Bigl\{\overline{\widehat{P}_I}\Bigr\}$ is the canonical conjugate of a variable $\overline{\widehat{{Q}}_j^{(i)}}$.
The variables $\Bigl\{\overline{\widehat{Q}_j^{(i)}}\Bigr\}$ play the role of the {\it pointer's coordinates} in a measurement apparatus (that is, the sensory organ) with respect to $\widehat{{\cal O}}$.
Each $\overline{{{Q}}_j^{(i)}}$ characterizes the state of the sensory cell system as a canonical variable.

In the following, we define the pointer's coordinate $\overline{\widehat{Q}_j^{(i)}}$ in a sensory cell.

Sensory organs have layers of cells for the processing of sensory transduction.
We denote by $\ell(\ge 1)$ the number of layers of the sensory organ in question.
In the $i$-th layer ($i=1,\ldots,\ell$), we assume there are $N^{(i)}$ sensory cells.
The sensory transduction process in the $i$-th layer is simplified to be within a definite time interval, $t^{(i)}_{\rm ini}\le t\le t^{(i)}_{\rm fin}$.

A huge number of sensory receptor cells are present in the first layer of sensory organs: $N^{(1)}\gg 1$.
In particular, in an ear, there are
\begin{equation}
\sim 1.5\cdot 10^4\ \ {\rm hair\ cells}\;,
\end{equation}
which are the auditory sensory receptor cells; in the retina of an eye, there are
\begin{equation}
\sim 4.5\cdot 10^6\ \ {\rm cone\ cells}\ \ {\rm and}\ \ \sim 9\cdot 10^7\ \ {\rm rod\ cells}\;,
\end{equation}
which are the visual sensory receptor cells \cite{Hodgkin}.

In auditory sensory transduction, the process occurs in the first layer (namely, $\ell=1$) and is directly induced by the migration of potassium ions from outside (that is, from the {\it endolymph}) to inside the hair cells.

In contrast, in visual sensory transduction, the process in the first layer is complicated and is induced by the migration of more than one type of cation.

We simplify the sensory transduction in such a way that the migration of one type of cation into the sensory cells in the $i$-th layer ($i=1,\ldots,\ell$) induces the transduction into the $i+1$-th layer (the $(\ell+1)$-th layer is the afferent nerve).
In this simplified model, we recall the case of the auditory sensory transduction (in this case, potassium ions are referred to as {\it cations}).

Now, we introduce the cation's mass concentration field, $Q^{(i)}_j(x_j^{(i)})$, in the spatial region $v_j^{(i)}$ ($j=1,2,\ldots,N^{(i)}$), of each sensory cell, relevant to the sensory transduction and consider its quantum field operator $\widehat{Q}^{(i)}_j(x_j^{(i)})$.
This operator acts in the quantum state space of the cation system in this cell.

In each sensory cell, the membrane electric potential has two types of {\it analog} change from the resting membrane electric potential---depolarization and hyperpolarization---as the consequence of changes in $Q^{(i)}_j(x_j^{(i)})$ induced by external stimuli.
Depolarization of the membrane electric potential gives rise to the release of neurotransmitters at the terminal of the corresponding sensory cell and subsequently excites the $(i+1)$-th layer \cite{Hodgkin}.

From this fact, we define the variables $\Bigl\{\overline{\widehat{{Q}}_j^{(i)}}\Bigr\}$ by
\begin{eqnarray}
\overline{\widehat{Q}_j^{(i)}}\equiv \frac{1}{v_j^{(i)}}\int_{v_j^{(i)}}\widehat{{Q}}_j^{(i)}(x_j^{(i)})d^3 x_j^{(i)}\;,\ \ j=1,2,\ldots,N^{(i)}\;,\ \ i=1,\ldots,\ell\;.\label{eq:Qdef}
\end{eqnarray}

From here, we further simplify the sensory transduction in two ways.
First, we ignore the non-linear processing part that occurs between two adjacent layers and at the $\ell$-th layer in the sensory organs.
(Particularly for visual sensory transduction, this part is complicated.)
Second, we ignore any distinction between sensory cells in the same layer (such as, between cone and rod cells in the first layer of the visual sensory organ).
Namely, we model a sensory organ as a {\it linear} filter for stimuli, and $N^{(i)}$ is common to all $i$.
This simplification is helpful in the present investigation because our aim is to examine the quantum measurement mechanism.

The canonical conjugate $P^{(i)}_j$ of $Q^{(i)}_j$ is defined by the negative of the cations's velocity potential \cite{KT1,KT2,KT3}.
Its quantum field operator $\widehat{P}^{(i)}_j(x^{(i)}_j)$ satisfies the canonical commutation relation
\begin{equation}
[\widehat{{Q}}^{(i)}_k(x^{(i)}_k),\widehat{P}^{(j)}_l(y^{(j)}_l)]={\rm i}\hbar\delta_{ij}\delta_{kl}\delta(x^{(i)}_k-y^{(j)}_l)\;.
\end{equation}
Due to the quantum mechanical {\it macroscopicity} of the sensory cells, the observables $\Bigl\{\overline{\widehat{P}^{(i)}_j}\Bigr\}$ $\Bigl(\overline{\widehat{P}^{(i)}_j}\equiv \int_{v_j^{(i)}} \widehat{P}^{(i)}_j$ $(x^{(i)}_j)d^3x_j^{(i)}\Bigr)$ are the canonical conjugate of variables $\Bigl\{\overline{\widehat{{Q}}^{(i)}_j}\Bigr\}$ that cannot be sharply measured, and are regarded as continuous superselection rule observables to a good approximation \cite{Neumann}.

\subsubsection{Mathematical model: Hamiltonians}

Now, the kinetic Hamiltonian of a sensory organ is
\begin{eqnarray}
\widehat{{H}}_{{\rm kin}}=\sum_{i=1}^\ell\sum_{j=1}^{N^{(i)}}\int_{v_j^{(i)}} \frac{1}{2}\nabla^{(i)}_j \widehat{P}_j^{(i)}(x_j^{(i)}) \cdot \bigl(\widehat{Q}_j^{(i)}(x_j^{(i)})\nabla^{(i)}_j \widehat{P}_j^{(i)}(x^{(i)}_j)\bigr)d^3x^{(i)}_j\;,\label{eq:Hso}
\end{eqnarray}
where we ignore the vorticity in the fluid motion of the cluster of cations solvated by water \cite{KT1,KT2,KT3}.

Here, it is worth quantifying the diffusion of cations in the sensory cells.
We consider the auditory case, where {\it cation} refers to a potassium ion in a hair cell.
The diffusion constant for a potassium ion in squid axoplasm is $D\approx 1.3\cdot 10^{-9}$ [m$^2\cdot$s$^{-1}$] \cite{Diff1,Diff2} and we use it.
From Fick's second law of three-dimensional diffusion, the diffusion distance $x_t$ for elapsed time $t$ is
\begin{eqnarray}
x_t&=&\sqrt{6Dt}\label{eq:xt}\\
   &=&c_x\cdot \sqrt{t}\;,\\
c_x&\approx&88\ [\mu{\rm m}\cdot {\rm s}^{-1/2}]\;.\label{eq:xt2}
\end{eqnarray}
In particular, to diffuse over a distance of $x_t=0.5$ [$\mu$m], it takes $t=x_t^2/(6D)\approx 3.2\cdot 10^{-5}$ [s].
This is the time scale of the auditory {\it response latency}, that is, the delay between a stimulus input and the onset of receptor current \cite{Diff2}.

To simplify the subsequent analysis, we reduce the quantum field operators $\overline{\widehat{Q}_j^{(i)}}$ and $\overline{\widehat{P}_j^{(i)}}$ to the quantum mechanical variables
\begin{eqnarray}
\overline{\widehat{Q}_j^{(i)}}&\overset{\rm reduce}{\longrightarrow}&\sum_{\overline{Q_j^{(i)}}}\overline{Q_j^{(i)}}\Bigl|\overline{Q_j^{(i)}}\Bigr>\Bigl<\overline{Q_j^{(i)}}\Bigr|\;,\\
\overline{\widehat{P}_j^{(i)}}&\overset{\rm reduce}{\longrightarrow}&\sum_{\overline{P_j^{(i)}}}\overline{P_j^{(i)}}\Bigl|\overline{P_j^{(i)}}\Bigr>\Bigl<\overline{P_j^{(i)}}\Bigr|\;,
\end{eqnarray}
respectively.
Here, $\overline{Q_j^{(i)}}$ and $\overline{P_j^{(i)}}$ are the spatially averaged eigenvalues.
Then, their eigenstates $\Bigl\{\Bigl|\overline{Q_j^{(i)}}\Bigr>\Bigr\}$ and $\Bigl\{\Bigl|\overline{P_j^{(i)}}\Bigr>\Bigr\}$ are in the restricted state space and are regarded as quantum mechanical eigenstates.

After this procedure, we adopt $\overline{\widehat{Q}^{(i)}_j}$ and $\overline{\widehat{P}^{(i)}_j}$ as the reduced quantum mechanical canonical variables of sensory cell $v_j^{(i)}$.

Since we treat the sensory organ as a linear filter for stimuli, there is a von Neumann-type interaction \cite{Neumann} between the sensory cells $\bigl\{v_j^{(i)}\bigr\}$ and their stimuli $\bigl\{{\cal O}^{(j)}\bigr\}$ with Hamiltonian
\begin{equation}
\widehat{{H}}_{\rm v.N.}=\sum_{i=1}^\ell\sum_{j=1}^{N^{(i)}}\Lambda_t^{(i)}{\cal E}_j\bigl(\widehat{{\cal O}}^{(j)}\bigr)\otimes \overline{\widehat{P}^{(i)}_j}\;.\label{eq:HVN}
\end{equation}
This models the diffusion process of cations from the outer cation reservoir into each sensory cell by the opening of the cation channels or the cation gates of the sensory cell.
For the sake of simplicity, each $\overline{P^{(i)}_j}$ of cations is assumed not to be changed by this diffusion process.

In Eq. (\ref{eq:HVN}), two types of function are introduced.
First, functions of time $\bigl\{\Lambda_t^{(i)}\bigr\}$, which satisfy
\begin{equation}
\Lambda_t^{(i^\prime)}=\delta_{ii^\prime}\Lambda^{(i)}\;,\ \ i^\prime=1,\ldots,\ell
\end{equation}
during $t^{(i)}_{\rm ini}\le t\le t^{(i)}_{\rm fin}$, are introduced.
Here, $\bigl\{\Lambda^{(i)}\bigr\}$ are time-independent positive constants.
Second, functions $\bigl\{{\cal E}_j\bigl({\cal O}^{(j)}\bigr)\bigr\}$ are introduced.
Each of these translates external stimulus ${\cal O}^{(j)}$ into an energy input ${\cal E}_j\bigl({\cal O}^{(j)}\bigr)$ (s.t., ${\cal E}_j(0)=0$) for sensory receptor cell $v_j^{(1)}$.

Then, before we apply the continuous superselection rule of the observables (that is, Eq. (\ref{eq:A8}) in ``Appendix A''),
\begin{equation}
\exp\Biggl[-\frac{{\rm i}}{\hbar}\int_{t^{(1)}_{\rm ini}}^{t^{(\ell)}_{\rm fin}} \widehat{H}_{\rm v.N.}dt\Biggr]\bigl|\bigl\{{\cal O}^{(j)}\bigr\}\bigr>\Bigl|\Bigl\{\overline{Q_j^{(i)}}\Bigr\}\Bigr>=\bigl|\bigl\{{{\cal O}}^{(j)}\bigr\}\bigr>\Bigl|\Bigl\{\overline{Q_j^{(i)}}+\Lambda^{(i)}{\cal E}_j\bigl({{{\cal O}}}^{(j)}\bigr)\delta t^{(i)}\Bigr\}\Bigr>
\end{equation}
holds in the sensory cells for $\delta t^{(i)}\equiv t^{(i)}_{\rm fin}-t^{(i)}_{\rm ini}$ ($i=1,\ldots,\ell$).\footnote{Here, we denote a simultaneous eigenstate, such as $|X^1,X^2,\ldots,X^l\ra$, by $|\{X\}\ra$.}
Namely, we obtain the relations
\begin{equation}
\delta \overline{Q^{(i)}_j}=\Lambda^{(i)}{\cal E}_j\bigl({{\cal O}}^{(j)}\bigr)\delta t^{(i)}\;,\ \ j=1,2,\ldots,N^{(i)}\;,\ \ i=1,\ldots,\ell\;.\label{eq:deltaQ}
\end{equation}
Here, $\delta\overline{Q^{(i)}_j}$ is the change of $\overline{Q^{(i)}_j}$ during $t^{(i)}_{\rm ini}\le t\le t^{(i)}_{\rm fin}$.

\subsubsection{Decoherence criterion}

Now, we consider the quantum mechanical uncertainties of $\Bigl\{\overline{Q^{(i)}_j}\Bigr\}$ and $\Bigl\{\overline{P^{(i)}_j}\Bigr\}$ and model these to be common to all sensory cells $\bigl\{v_j^{(i)}\bigr\}$ in the $i$-th layer ($i=1,\ldots,\ell$).
We denote these by $\Delta Q^{(i)}_0$ and $\Delta P^{(i)}_0$, respectively.

By using these uncertainties and invoking Eq. (\ref{eq:deltaQ}), we obtain the decoherence criterion on a dimensionless quantity that is the degree of progress of the decoherence mechanism due to the continuous superselection rule (see ``Appendix A'')
\begin{eqnarray}
\prod_{i=1}^\ell \prod_{j=1}^{N^{(i)}}\frac{\Bigl|\delta^2\overline{ Q_j^{(i)}}\Delta P_0^{(i)}/\hbar\Bigr|}{\Bigl|\sin (\delta^2\overline{ Q_j^{(i)}} \Delta P_0^{(i)}/\hbar)\Bigr|}\sim \prod_{i=1}^\ell \prod_{j=1}^{N^{(i)}}\frac{\Bigl|{\delta^2\overline{ Q_j^{(i)}}}/{\Delta Q_0^{(i)}}\Bigr|}{\Bigl|\sin (\delta^2\overline{ Q_j^{(i)}}/\Delta Q_0^{(i)})\Bigr|} \gg1\label{eq:Progress}
\end{eqnarray}
in the case of a single stimulus, $\widehat{{\cal O}}$, whose eigenstate takes the form
\begin{eqnarray}
|{\cal O}\ra\equiv|\overbrace{{\cal O}^{(\sigma(1))},\ldots,{\cal O}^{(\sigma(r))}}^{{\cal O}},\overbrace{{\cal O}^{(\sigma(r+1))},\ldots,{\cal O}^{(\sigma(N^{(1)}))}}^{0}\ra
\end{eqnarray}
with a definite permutation $\sigma$ and a definite natural number $r$.
(Note that ${\cal E}_j(0)=0$ for all $j$.)

Note that $\overline{Q^{(i)}_j}$ has the representation $m^{(i)}n^{(i)}_j/v_j^{(i)}$ where the mass of a cation is $m^{(i)}$ and the number of cations in $v_j^{(i)}$ is $n^{(i)}_j$.
So, Eq. (\ref{eq:Progress}) has the clearer expression 
\begin{equation}
\prod_{i=1}^\ell\prod_{j=1}^{N^{(i)}}\frac{\Bigl|\delta^2 n^{(i)}_j/{\Delta n^{(i)}_0}\Bigr|}{\Bigl|\sin (\delta^2 n^{(i)}_j/\Delta n^{(i)}_0)\Bigr|}\gg 1\;.\label{eq:gg1}
\end{equation}
In Eqs. (\ref{eq:Progress}) and (\ref{eq:gg1}), the product is taken over the sensory cells which transduce the stimulus ${\cal O}$.
Equation (\ref{eq:gg1}) is a kind of Bohr's criterion $n/\Delta n\gg 1$ on a classical mechanical object \cite{Umezawa}.

Now, this criterion is applied to the sensory organ as a combined system of the sensory cells that are considered as macroscopic bags of cation solution.

We consider the case of auditory sensory transduction.

In the process, a massive influx of potassium ions from the endolymph to the inside of the hair cells occurs.
Here, the high concentration of potassium ions, $\overline{Q}_{\rm end}$, in the endolymph is about one decimolar: $\overline{Q}_{\rm end}\sim 10^{-1}$ [mol$\cdot$m$^{-3}$]$\sim 10^5$ [$\mu$m$^{-3}$] \cite{Nature2}.
By using this, a small influx amount of potassium ions, $\delta n_0^{(1)}$, from an outer volume given just by the cubic size $v_{\rm out}\sim 10^{-2}$ [$\mu$m$^3$] of the tip link, whose length is $\sim 2\cdot 10^{-1}$ [$\mu$m] \cite{Tip}, into a hair cell is estimated to be $\delta n_0^{(1)}\sim 10^3$.
For a massive influx of potassium ions, $\delta n_j^{(1)}\gg \delta n_0^{(1)}$ holds.
Of course, $\delta n_j^{(1)}$ varies in accordance with the stimulus.

In the normal state (i.e., the incoherent matter state) of the potassium ions in $v_j^{(1)}\ll 10^4$ [$\mu$m$^3$]\footnote{The diameter and the depth of a hair cell are on the scales of a ten and a hundred of micrometers, respectively.
For this, see Fig. 13 in Ref. \cite{Hair1}.}, $\Delta n_0^{(1)}\ll \sqrt{n_j^{(1)}}$ holds.
The lower concentration of potassium ions in the hair cell is also about one decimolar \cite{Hair2}: $n_j^{(1)}$ is estimated to be $\overline{Q}_{\rm end}\cdot v_j^{(1)}\ll 10^9$.

These estimations support condition (\ref{eq:gg1}) in the case of auditory sensory transduction.

\subsection{Quantum measurement scheme involving sensory organs}

Now, we model selective quantum measurement of an external stimulus ${{\cal O}}$ via the sensory organs.

This consists of three steps.
In step 1, the term {\it non-selective measurement} of a pure state refers to a measurement step for which the resultant state is an exclusive statistical mixture of eigenstates of the observable in question, with the weights given by the Born rule: this step is familiarly known as the {\it decoherence process}.
(In the equations, the right arrow indicates the change of the density matrix according to the corresponding process.)

\begin{enumerate}
\item[Step 1] The first step is {\it non-selective measurement} of $\widehat{{\cal O}}$ due to the von Neumann-type interaction between the stimulus states $|{\cal O}\ra$ and the quantum state $|{\mathfrak A}_0\ra$ of the macroscopic sensory organs, assuming the existence of continuous superselection rule observables $\Bigl\{\overline{\widehat{P}_I}\Bigr\}$ of the sensory cells, within the time interval $t^{(1)}_{\rm ini}\le t\le t^{(\ell)}_{\rm fin}$:\footnote{It is a known fact that two state vectors belonging to different eigenspaces of $\widehat{{\cal O}}$ do not interfere at all with respect to the expectation values of all selected observables of the combined system $S$ (see Eq. (\ref{eq:A8})) after a time-dependent process of this von Neumann-type interaction \cite{Araki1,Ozawa}.
For details, see ``Appendix A''.}
\begin{eqnarray}
\widehat{\varrho}_{br}=\Biggl(\sum_nc_n|{\cal O}_n,{\mathfrak A}_0\ra\Biggr)|\{{\cal F}\}_0\ra\la \{{\cal F}\}_{0}|\Biggl(\sum_{n^\prime}\bar{c}_{n^\prime}\la{\cal O}_{n^\prime},{\mathfrak A}_0|\Biggr)
{\longrightarrow}\sum_n|c_n|^2|{\cal O}_n,{\mathfrak A}_0\ra|\{{\cal F}\}_0\ra\la \{{\cal F}\}_0|\la{\cal O}_n,{\mathfrak A}_0|\;.\label{eq:CSSR}
\end{eqnarray}

\item[Step 2] The second step is causal and continuous changes, according to the von Neumann equation of the density matrix, caused by an {\it entangling interaction} of the quantum feedback process occurring between the stimulus states $|{\cal O}\ra$ and the neural firing states $|\{{\cal F}\}\ra$.

This step is the result of the conversion of the changes of the value of ${Q}$, due to the external stimulus in step 1, into a neural firing pattern (that is, classical information) by the sensory organs:
\begin{eqnarray}
\widehat{\varrho}_{br}=
\sum_n|c_n|^2|{\cal O}_n,{\mathfrak A}_0\ra|\{{\cal F}\}_0\ra\la \{{\cal F}\}_0|\la {\cal O}_{n},{\mathfrak A}_0|
{\longrightarrow}
\sum_n|c_n|^2|{\cal O}_n,{\mathfrak A}_0\ra|\{{\cal F}\}_n\ra\la \{{\cal F}\}_{n}|\la{\cal O}_{n},{\mathfrak A}_0|\;.
\end{eqnarray}

The steps so far are the {\it objective} processes.

\item[Step 3] The third step is the {\it subjective} {\it event reading} subsequent to the objective steps, that is, the non-selective measurement supposing the state-reduction mechanism due to the existence of a coherence domain: \cite{arXiv2}
\begin{eqnarray}
\widehat{\varrho}_{br}=\sum_n|c_n|^2|{\cal O}_n,{\mathfrak A}_0\ra|\{{\cal F}\}_n\ra\la \{{\cal F}\}_n|\la{\cal O}_n,{\mathfrak A}_0|
\longrightarrow|{\cal O}_{n_0},{\mathfrak A}_0\ra|\{{\cal F}\}_{n_0}\ra\la \{{\cal F}\}_{n_0}|\la {\cal O}_{n_0},{\mathfrak A}_0|\;.\label{eq:dec2}
\end{eqnarray}
Here, as shown in Sect. 2.2, the system $\{{\cal F}\}$ consists of coherence domains of radiated photons \cite{PLA1,arXiv1}.

\end{enumerate}

\begin{figure}[htbp]
\begin{center}
\includegraphics[width=0.6\hsize,bb=0 0 260 93]{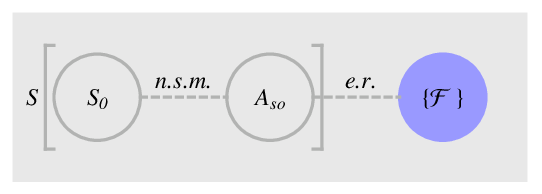}
\end{center}
\caption{
The proposed quantum measurement scheme in the human brain is schematically shown.
The measured system $S$, which consists of the system of an external stimulus $S_0$ and the sensory organ $A_{so}$ as a macroscopic measurement apparatus, undergoes non-selective measurement (n.s.m.) due to the continuous superselection rule in $S$.
After the non-selective measurement, an event reading (e.r.) is done by the measuring system, with a neural firing pattern $\{{\cal F}\}$, that consists of coherence domains.}
\end{figure}

This measurement scheme of the human brain involving sensory organs is a {\it type I selective measurement} in the classification of the selective measurements proposed in Ref. \cite{JSTAT}.
Namely, the measuring system $\{{\cal F}\}$ that reads measurement events is independent of the measured system $S$ (see Fig. 1) on which the non-selective measurement process (step 1) acts.

Therefore, due to the result in Ref. \cite{JSTAT}, an event reading in the third process requires internal work (i.e., an energy transfer) from the measuring system $\{{\cal F}\}$ to the measured system $S$ of an amount $k_BT$ for the Boltzmann constant $k_B$ and the temperature $T$ of the system $S$: that is, the event reading in the selective quantum measurement scheme {\it via the sensory organs} $A_{so}$ is a {\it physical} process.
This agrees with our experience.

Our quantum measurement scheme requires a macroscopically coherent quantum description of {\it dynamical} neural firing states that is compatible with our classical coding process.
Here, {\it our classical coding process} is defined by using neural firing and resting states as one classical bit for each neuron.

However, the quantum models of the brain proposed so far (e.g., Refs. \cite{Vitiello,PH1,PH2}) describe physical processes common to both neural firing and resting states of all neurons; thus, these processes are not thought to be compatible with our classical coding process.\footnote{These quantum models are thought to be compatible with their respective classical coding processes, although in ways different from ours.
For details of the models proposed in Ref. \cite{Vitiello} and Refs. \cite{PH1,PH2}, see Ref. \cite{FV} and Ref. \cite{PH3}, respectively.}

In contrast, due to our result, the quantum coherence structure of our quantum state of a neuron emerges only when an action potential is generated.
Therefore, in a time-dependent process coupled over the whole neural network, it enables us to describe each dynamical neural firing state by a macroscopically coherent quantum state.
This quantum state {\it is} compatible with our classical coding process.

Next, we overview Ref. \cite{arXiv2}.

In Ref. \cite{arXiv2}, the author proposed a mechanism for event reading, with respect to energy, occurring in an arbitrary {\it coherence domain} due to the time-increment fluctuation attributed to spontaneous time reparametrization symmetry breaking in canonical gravity.

This mechanism works in the framework of an extended model of canonical gravity theory, under two assumptions:
\begin{enumerate}
\item[A1] The hypothesis of the existence of one additional massive spin-$1/2$ particle as cold dark matter having a long-range self-interaction and an exchange interaction with the spinor expression of a particular type of the temporal part of the space-time metric (the coordinate system).

\item[A2] A novel interpretation of quantum mechanics with respect to the quantum mechanical position uncertainties of particles, where we reverse the roles of particles and the coordinate system in the uncertainties of the positions of particles.
\end{enumerate}

In this model, the additional particles play the role of creating the time reparametrization invariant potential, and the {\it projection hypothesis} (that is, step 3) is physical and is realized as the {\it system of the additional particles confined in a coherence domain}.

Finally, we conclude this section: the numerical results (\ref{eq:NUM1}) and (\ref{eq:NUM2}) in Ref. \cite{arXiv1} offer a possible explanation for the role played by the human brain as a quantum measurement apparatus when we suppose a state-reduction mechanism in step 3 {\it in the coherence domain} \cite{arXiv2}.

\section{The second model: the resonant QED nature}
This section consists of two parts.

In Sect. 3.1, we briefly explain the fundamental properties of {\it coherence domains}.
This will be the minimum needed to account for the setup in Sect. 3.2.

In Sect. 3.2, we will find that a quantum measurement scheme resembling the scheme of the first resonant model exists in the generic system of an assembly of Preparata
et al.'s
coherence domains.
(Note that the first resonant model is in coherence domains, with the dimension $l_c$ and not related to a superradiant phase transition.)

The first resonant model is induced by dynamical instability, as in the FEL model, but the second resonant model is induced by a superradiant phase transition.

\subsection{Preliminaries}
\subsubsection{Coherence domain}

In QED with {\it resonant} interactions between matter (throughout this section, we assume that matter can be approximately described via a quantum mechanical {\it two-level} system) and radiation, there exists a {\it non-perturbative} coherent ground state if the effective coupling constant $q\sqrt{n}$, enlarged by the factor $\sqrt{n}$ in the rescaled action of the fields rescaled by the factor $1/\sqrt{n}$\footnote{The origin of this factor $\sqrt{n}$ in the effective coupling constant is the fact that the resonant coupling is a triplet of the rescaled fields but the other terms in the action are doublets of the rescaled fields.} \cite{Preparata}, for the electric charge $q$ and the number density $\rho=n/V$ of {\it quasi-particles} (which define the ground state of the system as their vacuum state) within a domain having a volume $V$ exceeds a threshold $q\sqrt{\rho_cV}$ that depends on the electric polarizability of the quasi-particles\footnote{Specifically, as the limit cycle of the system approaches the completely {\it inverse population} (a term used in laser physics), the critical quasi-particle number grows monotonically to infinity.
This is in contrast with the situation for a laser.
The system is regarded as a kind of laser phenomenon whose realization requires no pumping process.}.
Furthermore, a superradiant phase transition occurs if the temperature $T$ is below a critical value $T_c(\rho)$ to avoid boil-off of this domain due to thermal fluctuations.
In summary, the conditions for a superradiant phase transition are
\begin{eqnarray}
\rho&>& \rho_c\;,\label{eq:Condition1}\\
T&<&T_c(\rho)\;.\label{eq:Condition2}
\end{eqnarray}

\begin{figure}[htbp]
\begin{center}
\includegraphics[width=0.5 \hsize,bb=0 0 260 197]{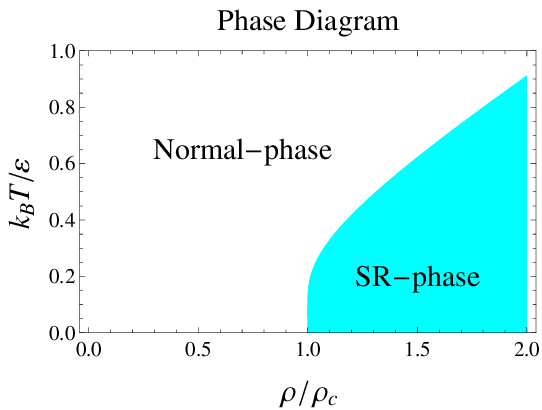}
\end{center}
\caption{For the quantum Dicke--Preparata model, the superradiant phase (SR-phase) and the normal phase are as shown in the phase diagram in the $(\rho,T)$-plane.
Here, $\varepsilon$ is the energy gap in the two-level approximation.
This figure appears as Fig. 2 in Ref. \cite{SWS}.}
\end{figure}

For the {\it Dicke--Preparata model} in which radiation is treated quantum mechanically as a single resonant photon oscillator mode $k_0$ \cite{SWS}, the phase diagram is as shown in Fig. 2.
In this model, the critical quasi-particle number density $\rho_c$ is given by
\begin{equation}
\rho_c=\frac{2\epsilon_0\varepsilon}{(\varepsilon_{k_0}\cdot d_{10})^2}\;,
\end{equation}
where the vacuum permittivity is $\epsilon_0$, the photon polarization vector is $\varepsilon_{k_0}$, $d_{10}$ is the excitation matrix element of the electric dipole moment operator vector of the quasi-particle, and the critical temperature $T_c(\rho)$ for $\rho>\rho_c$ is given by \cite{SWS}
\begin{equation}
T_c(\rho)=\frac{\varepsilon}{k_B\ln ((\rho+\rho_c)/(\rho-\rho_c))}\;.
\end{equation}

This non-perturbative coherent ground state is a solution of the equations of motion and three conservation laws (conserving the total number of quasi-particles, the total momentum, and the total energy of the system).
This solution's electromagnetic field amplitude has evolved by running away from the solution in the gas-like perturbative QED ground state.

In the non-perturbative ground state, bosons (not quasi-particles) are condensed with energy ${\cal E}<{\cal E}_0$, where ${\cal E}_0$ is the energy of the perturbative ground state \cite{Preparata,Enz}.
Due to this fact, after the superradiant phase transition, to increase the energy gain ${\cal E}_0-{\cal E}>0$, the quasi-particle system is bound to assume the highest possible density \cite{Many1}.

\begin{figure}[htbp]
\begin{center}
\includegraphics[width=0.5 \hsize,bb=0 0 260 258]{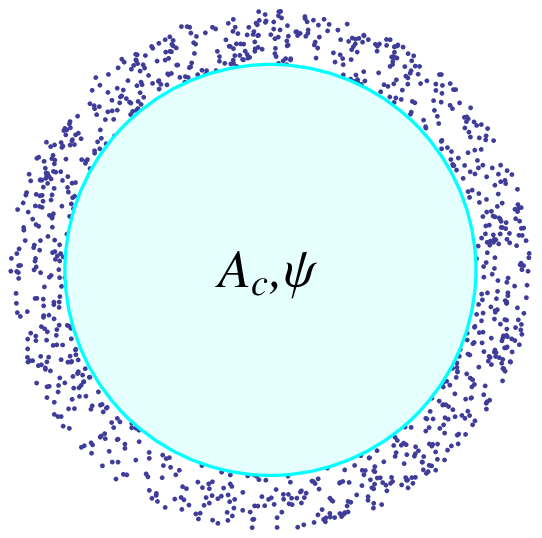}
\end{center}
\caption{
Schematic planar projection of a coherence domain.
Total reflection of the electromagnetic field occurs and electromagnetic coherence is realized within the cyan domain.
The surrounding dots represents the domain where ${\rm grad}{A}_c\neq 0$ holds for the classical electromagnetic field $A_c$.
$\psi$ denotes the configuration of the quasi-particles.
Within this dotted domain, the electromagnetic field $A_c$ is evanescent and decays exponentially.}
\end{figure}

A superradiant phase transition occurs in a domain with the spatial scale of the resonance wavelength of radiated photons (i.e., the coherence length).
In this {\it coherence domain}, all quasi-particles are coupled to the electromagnetic field in the same way: their time evolution is homogeneous.
Coherence domains were introduced in Ref. \cite{Preparata} and earlier research along that line (see Fig. 3).

In the limit cycle of the system giving rise to the non-perturbative ground state, as an ansatz, inside a coherence domain, the matter (atoms or molecules of the same type) system oscillates between the two energy-level states in phase, and the coherent electromagnetic field oscillates too.
Then, the frequency of the coherent electromagnetic field gets shifted to a smaller value by interaction with the matter.
Owing to such a negative shift of frequency, the coherent electromagnetic field is kept trapped inside the coherence domain by a mechanism completely analogous to the well-known {\it total reflection} that is experienced by light at the interface between two media of different refraction indices (see the cyan domain in Fig. 3), and this negative shift plays the role of a spontaneously created cavity in laser physics \cite{Preparata,GP}.

As shown in Fig. 3, an evanescent electromagnetic wave with an effective photon mass appears, as a result of the quadruple matter-photon interaction in QED, because of the homogeneity (i.e., the slow space variation) of the density function of quasi-particles in the coherence domain.
This evanescent electromagnetic wave pulsates with the shifted frequency and dies off at a rate on the order of its wavelength, that is, the dimension of the coherence domain (see the dotted domain in Fig. 3).
Namely, the coherence domain emits no photons, in contrast with a conventional laser system \cite{Preparata,GP}.

When different coherence domains overlap with each other through their tails, inside which evanescent electromagnetic fields are trapped, these coherence domains are mutually attracted in order to decrease the total energy of the system of coherence domains.
More generally, in an ensemble of many coherence domains, when the component coherence domains come as close as possible, the energy of the ensemble is minimized \cite{Many1}.

Note that the superradiant phase transition is a {\it spontaneous process} (i.e., one with no cavity and no pump).

As an illustration of Preparata
et al.'s
theory of superradiant phase transitions, under three approximations (the electric dipole approximation, the two-level approximation and the rotating-wave approximation), previous theoretical calculations have shown that pure liquid water is a two-phase system of water molecules in a coherent superradiant state and the incoherent normal state.
In this case, the two phases {\it coexist} exactly like in Landau's two-fluid theory of superfluid $^4$He \cite{KT1,KT2,KT3}, and their fractions in pure liquid water depend on the temperature \cite{Many1,GP,Many2}.

In this picture of pure liquid water, the coherence domain for the transitions between the electronic ground and first excited states (here, not the rotational ground and first excited states) of water molecules has a radius of $375$ [\AA] at $0$ [K], and this decreases to $250$ [\AA] at room temperature ($300$ [K]) \cite{Many2}.
The fraction of the coherent state is $0.28$ at room temperature \cite{Many2}.

\subsubsection{The resonant system: Hamiltonians}

In the following discussion, we consider a general Preparata
et al.'s
superradiant system of $n$ atoms or molecules of the same type (we call these {\it elements}) and a resonant radiation field, using the electric dipole approximation of each element within the coherence domain and using the two-level approximation of each element.

The Hamlitonian of the system of radiation and elements consists of three terms:
\begin{equation}
\widehat{H}=\widehat{H}_{\rm em}+\widehat{H}_{\rm el}+\widehat{H}_{\rm int}\;.
\end{equation}

In the following, we explain each of the three terms of $\widehat{H}$ in the radiation gauge.

The first term, $\widehat{H}_{\rm em}$, is the electromagnetic free Hamiltonian in the coherence domain $V$:
\begin{equation}
\widehat{H}_{\rm em}=\frac{1}{2}\int_{V}\biggl\{\epsilon_0|{\widehat{E}}({x},t)|^2+\frac{1}{\mu_0}|\nabla \times {\widehat{A}}({x},t)|^2\biggr\}d^3x
\end{equation}
for the electric field vector operators ${\widehat{E}}$ that satisfy the canonical commutation relations with the transverse $\epsilon_0{\widehat{A}}$.
Here, $\epsilon_0$ and $\mu_0$ are the vacuum permittivity and the vacuum permeability, respectively.

By using the mode expansions (in this, $k=({\bk},\lambda)$ and $\varepsilon_k$ denote a composite of wavenumber vector ${\bk}$ with polarization index $\lambda=\pm$ and a polarization vector, respectively\footnote{The polarization vectors $\varepsilon_k=\varepsilon_k^\ast$ satisfy $\varepsilon_k=\varepsilon_{-k}$ ($-k\equiv (-\bk,-\lambda)$), ${\bk}\cdot \varepsilon_k=0$ and $\varepsilon_{\bk,\lambda}\cdot \varepsilon_{\bk,\lambda^\prime}=\delta_{\lambda\lambda^\prime}$.})
\begin{eqnarray}
\widehat{A}(x,t)&=&\sum_k\sqrt{\frac{ \hbar }{\epsilon_0\omega_{\bk} V}}\widehat{q}_k{\varepsilon}_ke^{{\rm i}{\bk}\cdot x}\;,\\
\widehat{E}(x,t)&=&-\sum_k\sqrt{\frac{ \hbar \omega_{\bk}}{\epsilon_0V}}\widehat{p}_{-k}{\varepsilon}_ke^{{\rm i}{\bk}\cdot x}\;,
\end{eqnarray}
with $\omega_{\bk}=c|{\bk}|$, this term can be written as
\begin{equation}
\widehat{H}_{\rm em}=\frac{\hbar}{2}\sum_{k}\{\omega_{\bk}\widehat{p}_{-k}\widehat{p}_k+\omega_{\bk} \widehat{q}_k\widehat{q}_{-k}\}\;,
\end{equation}
where
\begin{equation}
[\widehat{q}_k,\widehat{p}_{k^\prime}]={\rm i}\delta_{kk^\prime}\;.
\end{equation}

Furthermore, by using the creation and annihilation operators for photons
\begin{eqnarray}
\widehat{a}_k&=&\frac{1}{\sqrt{2}}\{\widehat{q}_k+{\rm i}\widehat{p}_{-k}\}\;,\\
\widehat{a}_k^\dagger&=&\frac{1}{\sqrt{2}}\{\widehat{q}_{-k}-{\rm i}\widehat{p}_k\}\;,
\end{eqnarray}
which satisfy
\begin{equation}
[\widehat{a}_k,\widehat{a}_{k^\prime}^\dagger]=\delta_{kk^\prime}\;,
\end{equation}
we obtain another expression
\begin{equation}
\widehat{H}_{\rm em}=\sum_k\hbar\omega_k\widehat{a}_k^\dagger \widehat{a}_k\;.
\end{equation}

The second term, $\widehat{H}_{\rm el}$, is the free Hamiltonian of electric dipoles of elements ($i=1,2,\ldots,n$).

As we did in the first model, we describe the electric dipole moment operators of elements in terms of their respective energy spin operators $\widehat{s}^a$ ($a=1,2,3$).
These energy spin operators obey an $su(2)$ algebra, as in Eqs. (\ref{eq:Spin1}) to (\ref{eq:Spin3}), with energy spinors in the two-dimensional energy state space spanned by their associated ground state $|g\ra$ and their associated excited energy state $|e\ra$.

Then, due to this two-level approximation, $\widehat{H}_{\rm el}$ takes the form
\begin{equation}
\widehat{H}_{\rm el}={\varepsilon}_{\rm el}\sum_{i=1}^n\widehat{s}^3_i\label{eq:TLA}
\end{equation}
for an energy gap ${\varepsilon}_{\rm el}=\hbar \Omega$ between the two energy levels.

The third term, $\widehat{H}_{\rm int}$, is the interaction Hamiltonian for the resonant coupling between the radiated photons and the electric dipoles of elements \cite{Loudon}:
\begin{eqnarray}
\widehat{H}_{\rm int}&=&\frac{{\rm i}\hbar}{\sqrt{2}}\sum_{i=1}^n\sum_{k\in S_\Omega}\lambda_{k,i}\{\widehat{a}_k\widehat{s}_i^+e^{{\rm i}\bk\cdot x_i}-\widehat{a}^\dagger_k\widehat{s}_i^-e^{-{\rm i}\bk\cdot x_i}\}\label{eq:int1}\\
&=&-\hbar\sum_{i=1}^n\sum_{k\in S_\Omega}\lambda_{k,i}\{\widehat{q}_{-k}\widehat{s}^2_i+\widehat{p}_k\widehat{s}_i^1\}e^{-{\rm i}\bk\cdot x_i}\;,\label{eq:couplings}
\end{eqnarray}
where $S_\Omega=\{k|\omega_\bk=\Omega\}$.
Here, we introduce
\begin{eqnarray}
\widehat{s}^+&=&\widehat{s}^1+{\rm i}\widehat{s}^2\;,\\
\widehat{s}^-&=&\widehat{s}^1-{\rm i}\widehat{s}^2
\end{eqnarray}
and
\begin{eqnarray}
\lambda_{k,i}= -\sqrt{\frac{\omega_\bk}{\epsilon_0\hbar V}}{\varepsilon}_k\cdot d_{i,10}\;,\ \ k\in S_\Omega\;,\ \ i=1,2,\ldots,n
\end{eqnarray}
for the off-diagonal matrix element $d_{10}\equiv \la e|\widehat{d}|g\ra$ of the electric dipole moment operator vector $\widehat{d}$ of each element (matrix indices $1$ and $0$ refer to the states $|e\ra$ and $|g\ra$, respectively).
Here, we remember that, since the electric dipole moment operator has odd parity, it has no non-zero diagonal matrix elements: $d_{00}=d_{11}=0$.
In Eq. (\ref{eq:int1}), we impose the reality condition $d_{10}=d_{01}$ \cite{Enz,Loudon}, then we obtain $\widehat{d}=2d_{10}\widehat{s}^1$.

\subsubsection{The resonant system: spontaneous symmetry breaking}

The system $\widehat{H}$ has a proper $U(1)$ symmetry under the simultaneous global transformations parameterized by $0\le \theta<2\pi$
\begin{eqnarray}
\widehat{q}_k&\longrightarrow&\widehat{q}_k\cos \theta -\widehat{p}_{-k}\sin\theta\;,\\
\widehat{p}_{-k}&\longrightarrow&\widehat{q}_k\sin \theta+\widehat{p}_{-k}\cos\theta\;,\\
\widehat{s}_i^1&\longrightarrow&\widehat{s}_i^1\cos \theta+\widehat{s}_i^2\sin \theta\;,\\
\widehat{s}_i^2&\longrightarrow&-\widehat{s}_i^1\sin\theta+\widehat{s}_i^2\cos \theta\;,\\
\widehat{s}_i^3&\longrightarrow&\widehat{s}_i^3
\end{eqnarray}
for $i=1,2,\ldots,n$ and has non-perturbative ground states $|0(A_c,s_c)\ra$ that are infinitely degenerate with respect to this global $U(1)$ symmetry.

Each ground state $|0(A_c,s_c)\ra$ gives rise to time-independent classical vector fields ${A}_c({x})$ and ${s}_c$ as its vacuum expectation values:
\begin{eqnarray}
{A}_c({x})&=&\Bigl<0(A_c,s_c)\Bigl|{\widehat{A}}({x},t)\Bigr|0(A_c,s_c)\Bigr>\;,\\
{s}_c&=&\Bigl< 0(A_c,s_c)\Bigl|{\widehat{s}}({x},t)\Bigr|0(A_c,s_c)\Bigr>\;.\label{eq:Sc}
\end{eqnarray}
In the Heisenberg picture, these are the time-independent solution of the Heisenberg equations:
\begin{eqnarray}
\dot{\widehat{q}}_k&=&-\frac{{\rm i}}{\hbar}[\widehat{q}_k,\widehat{H}]=\Omega\widehat{p}_{-k}-\sum_{i=1}^n\lambda_{k,i}\widehat{s}_i^1e^{-{\rm i}\bk\cdot x_i}\;,\\
\dot{\widehat{p}}_{-k}&=&-\frac{{\rm i}}{\hbar}[\widehat{p}_{-k},\widehat{H}]=-\Omega \widehat{q}_k+\sum_{i=1}^n \lambda_{k,i} \widehat{s}_i^2e^{-{\rm i}\bk\cdot x_i}\;,\\
\dot{\widehat{s}}^1_i&=&-\frac{{\rm i}}{\hbar}[\widehat{s}_i^1,\widehat{H}]=-\Omega \widehat{s}_i^2-\sum_{k\in S_\Omega}\lambda_{k,i}\widehat{q}_{-k}\widehat{s}^3_ie^{-{\rm i}\bk\cdot x_i}\;,\label{eq:s1EOM}\\
\dot{\widehat{s}}^2_i&=&-\frac{{\rm i}}{\hbar}[\widehat{s}_i^2,\widehat{H}]=\Omega \widehat{s}_i^1+\sum_{k\in S_\Omega}\lambda_{k,i}\widehat{p}_k\widehat{s}_i^3e^{-{\rm i}\bk\cdot x_i}\;,\label{eq:s2EOM}\\
\dot{\widehat{s}}^3_i&=&-\frac{{\rm i}}{\hbar}[\widehat{s}_i^3,\widehat{H}]=\sum_{k\in S_\Omega}\lambda_{k,i}\{\widehat{q}_{-k}\widehat{s}_i^1-\widehat{p}_k\widehat{s}_i^2\}e^{-{\rm i}\bk\cdot x_i}\;,\label{eq:s3EOM}
\end{eqnarray}
where $k\in S_\Omega$ and $i=1,2,\ldots,n$.
The vacuum expectation values of all of lines are zero:
\begin{eqnarray}
0&=&\Omega \la \widehat{p}_{-k}\ra-\sum_{i=1}^n\lambda_{k,i}\la \widehat{s}_i^1\ra e^{-{\rm i}\bk\cdot x_i}\;,\\
0&=&-\Omega \la \widehat{q}_k\ra+\sum_{i=1}^n \lambda_{k,i} \la\widehat{s}_i^2\ra e^{-{\rm i}\bk\cdot x_i}\;,\\
0&=&-\Omega\la \widehat{s}_i^2\ra-\sum_{k\in S_\Omega}\lambda_{k,i}\la \widehat{q}_{-k}\ra\la \widehat{s}^3_i\ra e^{-{\rm i}\bk\cdot x_i}\;,\\
0&=&\Omega \la \widehat{s}_i^1\ra+\sum_{k\in S_\Omega}\lambda_{k,i}\la \widehat{p}_k\ra\la \widehat{s}_i^3\ra e^{-{\rm i}\bk\cdot x_i}\;,\\
0&=&\sum_{k\in S_\Omega}\lambda_{k,i}\{\la \widehat{q}_{-k}\ra\la\widehat{s}_i^1\ra -\la \widehat{p}_k\ra\la \widehat{s}_i^2\ra \}e^{-{\rm i}\bk\cdot x_i}\;.
\end{eqnarray}

The classical fields $A_c(x)$ and $s_c$---that is, the vacuum expectation values of the quantum field operators ${\widehat{A}}({x},t)$ and ${\widehat{s}}({x},t)$, respectively---satisfy
\begin{eqnarray}
A_c(x)&=&\sum_{k\in S_\Omega} \sqrt{\frac{\hbar}{\epsilon_0\Omega V}}\la \widehat{q}_k\ra{\varepsilon}_ke^{{\rm i}\bk\cdot x}\neq 0\;,\\
\la \widehat{q}_k\ra&=&\frac{1}{\Omega}v\sin\theta_0\sum_{i=1}^n \lambda_{k,i}e^{-{\rm i}\bk\cdot x_i}\;,\\
\la\widehat{p}_{-k}\ra&=&\frac{1}{\Omega}v\cos\theta_0\sum_{i=1}^n\lambda_{k,i} e^{-{\rm i}\bk\cdot x_i}\;,\\
s^1_c&=&v\cos\theta_0\;,\label{eq:sc1}\\
s^2_c&=&v\sin\theta_0\;,\label{eq:sc2}\\
s^3_{i,c}&=&-\Omega^2\Biggl[\sum_{k\in S_\Omega}\Biggl\{\lambda_{k,i}\sum_{j=1}^n \lambda_{k,j}e^{-{\rm i}\bk\cdot (x_i-x_j)}\Biggr\}\Biggr]^{-1}
\end{eqnarray}
for constants $v$ and $\theta_0$.
Here, the global $U(1)$ symmetry is spontaneously broken.

Hereafter, to simplify the argument, we consider the case of a system $\widehat{H}$ of radiation and rotating water molecules (elements).
A remarkable property of this system $\widehat{H}$ is that the ground state has a `ferromagnetically' ordered energy spin ground state part in which all energy spin directions align in one direction (see Eqs. (\ref{eq:sc1}) and (\ref{eq:sc2})).
Due to this property, when the system of elements is around a ferroelectric with an electric dipole moment ${{S}}^a_{\rm b.c.}$ that is directed strongly enough, the alignment in the energy spin ground state is in the direction of this boundary condition ${S}^a_{\rm b.c.}$ (refer to Fig. 4).

\begin{figure}[htbp]
\begin{center}
\includegraphics[width=0.5 \hsize,bb=0 0 260 176]{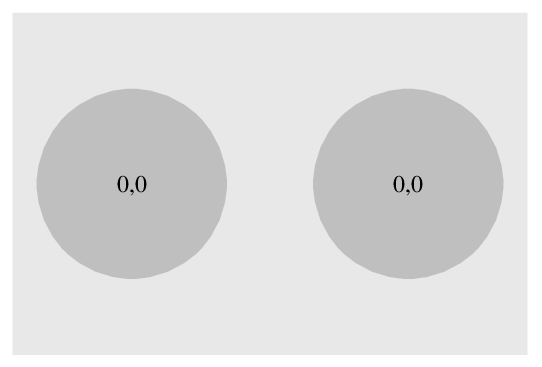}
\includegraphics[width=0.5 \hsize,bb=0 0 260 176]{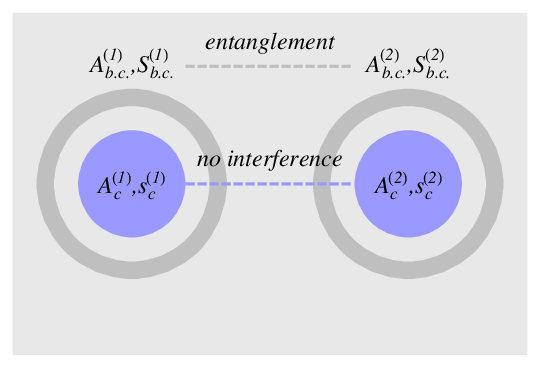}
\end{center}
\caption{
The formation of a non-perturbative ground state $|0(A_c,s_c)\ra$ for two coherence domains (two center blue domains in the lower panel) with their respective narrow-sense boundary conditions $(A_{\rm b.c.}, S_{\rm b.c.})$ is schematically shown.
In the upper panel, the global $U(1)$ symmetry is unbroken and there is no narrow-sense boundary condition.
In the lower panel, the global $U(1)$ symmetry is broken, the narrow-sense boundary conditions $(A_{\rm b.c.}^{(1)},S_{{\rm b.c.}}^{(1)})$ and $(A_{\rm b.c.}^{(2)},S_{{\rm b.c.}}^{(2)})$ are entangled, and the states of coherence domains with different vacuum expectation values $(A_c^{(1)},s_c^{(1)})$ and $(A_c^{(2)},s_c^{(2)})$ for the spontaneously broken global $U(1)$ symmetry have no interference (see Sect. 3.2 for an explanation).}
\end{figure}

A crucial feature of this energy spin system is that even if the boundary condition ${S}^a_{\rm b.c.}$ disappears, since the energy ${\cal E}$ of $|0(A_c,s_c)\ra$ is less than the perturbative ground state energy ${\cal E}_0$, the system will not spontaneously return to its state before the appearance of ${S}^a_{\rm b.c.}$.
Thus, the `memory' $s^a_c$, which is an extended object with a macroscopically `ferromagnetically' ordered energy spin configuration that spontaneously breaks the spatial rotational symmetry (SRS), is written {\it stably} in the non-perturbative ground state $|0(A_c,s_c)\ra$.
After the removal of the boundary condition ${S}^a_{\rm b.c.}$, when a weak perturbation on the boundary condition $\Delta{S}^a_{\rm b.c.}$ is added to the system, a gapless Goldstone mode of the spontaneously broken global SRS arises due to the Nambu-Goldstone theorem.
In the case of this system, the boson of this mode is the {\it polariton} (i.e., the low-energy excitation of the electric dipole system) \cite{Yasue}.\footnote{For a comprehensive and pedagogical review of Refs. \cite{Enz,STU2,JPY}, see Ref. \cite{Yasue}.}

Here, the Goldstone field is identified with a local (i.e., space-time dependent) $U(1)$ phase fluctuation variable $\theta(x,t)$ of the energy spinor field, $\psi(x,t)$, of the elements.
This Goldstone field restores the broken global $U(1)$ symmetry as a local $U(1)$ symmetry under the transformation
\begin{equation}
\psi(x,t)\longrightarrow \psi^\prime(x,t)=e^{i\theta(x,t)\sigma_3/2}\psi(x,t)\;.
\end{equation}
Such a local $U(1)$ symmetry can be ensured by the corresponding gauge transformation of the classical radiation field:
\begin{equation}
A(x,t)\longrightarrow A^\prime(x,t)=A(x,t)-\nabla \theta(x,t)\;.\label{eq:AH}
\end{equation}
Equation (\ref{eq:AH}) means that the classical radiation field $A^\prime(x,t)$ absorbs the Goldstone field $\theta(x,t)$ into its longitudinal wave component.
As a result, $A^\prime(x,t)$ obeys the Klein-Gordon equation with a photon mass term.
This is the Anderson-Higgs mechanism \cite{Umezawa,JPY}.

The account of the setup of the system has now been completed.

\subsection{Quantum measurement scheme}

With the setup described above, according to the quantum field theory of memory advocated by Umezawa et al. \cite{RU,STU1,STU2,Vitiello,JPY}, we combine the results from two different theories, Umezawa's theory of extended objects in quantum field theory \cite{Umezawa,JMP1,JMP2,PRD} and Preparata
et al.'s
theory of coherence domains in QED, to explain the universal memory function of {\it the resonant QED nature}.\footnote{In addition to the present study, Refs. \cite{Bio1,Bio2} use Umezawa et al.'s theory of extended objects as a universal formalism.
However, these works precede Preparata et al.'s theory of coherence domains in QED and do not discuss quantum measurement.}

To formulate this problem, we divide the QED world into two parts: coherence domains and their boundary conditions.
Formally, the structure of the state space ${\mathfrak H}_W$ of the QED world is
\begin{equation}
{\mathfrak H}_W=\Biggl(\bigotimes_{{\mathfrak D}_c}{\mathfrak H}_{CD}^{({{\mathfrak D}_c})}\Biggr)\otimes {\mathfrak H}_{\rm b.c.}
\end{equation}
for the state spaces ${\mathfrak H}_{CD}^{({{\mathfrak D}_c})}$ and ${\mathfrak H}_{\rm b.c.}$ of the coherence domains $\{{{\mathfrak D}_c}\}$ and their boundary conditions $\{{\cal B}\}$, respectively.
The sources of information to be transduced, in other words, the {\it measured objects} are the time-dependent boundary conditions $\{{\cal B}\}$.

The physical processes in the memory function can be schematically shown as
\begin{equation}
{\xymatrix{{\rm Coherence\ Domains} \ar@{>}@[black][rrr]^{{\rm superradiance}}&&&|0\ra\ar@{>}@[black][lll]\ar@/_10pt/@(ru,rd)[]|{{\rm{read}}^3}\\
&&&\\
{{\rm\ Ordered\ Patterns\ of \ Matter}}\ar[uu]|{{\rm living\ in}}\ar[rrruu]|{{\rm written^1/retrieved^2}}&&&}}\label{eq:diagram6}
\end{equation}
for a non-perturbative ground state $|0\ra=|0(A_c,s_c)\ra$ with a spontaneously broken symmetry, evanescent electromagnetic field $A_c$ and electric dipole field $s^a_c$.

The memory function of this absolutely {\it open} quantum system of coherence domains consists of three physical processes for the ground state $|0(A_c,s_c)\ra$: {\it writing}, {\it retrieval} and {\it reading}.

We explain these processes in four steps.
\begin{enumerate}
\item[Step 1] First, ordered patterns of extended objects are {\it written} by their boundary conditions $\{{\cal B}\}$ as stable {\it{memories}} in the ground state $|0(A_c,s_c)\ra$ with infinitely many varieties via the condensation mechanism of bosons in the ground state, that is, the spontaneous symmetry breaking.

This mechanism gives non-zero vacuum expectation values of quantum field operators $\widehat{\Psi}={\widehat{A}},{\widehat{s}}$:
\begin{eqnarray}
{\cal B}:|0;0\ra\longrightarrow|0;{\cal M}\ra\;,\ \ {\cal M}({x})=\Bigl< 0\Bigl|\widehat{\Psi}({x},t)\Bigr|0\Bigr>\;,
\end{eqnarray}
where $|0;0\ra$ and $|0;{\cal M}\ra$ indicate $|0(0,0)\ra$ and $|0(A_c,s_c)\ra$, respectively.
Here, we use the term {\it memory} for vacuum expectation values ${\cal M}(x)$ due to the direct integral structure of the Hilbert space of the system $\Psi$ over the subspaces labeled by ${\cal M}(x)$.
In the limit $\hbar\to 0$, ${\cal M}(x)$ obeys the same equations of motion as $\widehat{\Psi}(x,t)$ and describes the classical mechanical behavior of the extended object \cite{Umezawa,JMP1}.

The quantum field system of the extended object has degrees of freedom of three kinds: quasi-particles, quantum mechanical degrees of freedom $\widehat{q}$ and $\widehat{p}$ of the zero-energy translation mode of the extended object, and the classical mechanical order parameter field ${\cal M}$ \cite{Umezawa,JMP1,JMP2,PRD}.

The total Hilbert space ${\mathfrak H}_{CD}^{({\mathfrak D}_c)}$ of the quantum states $|\Psi;{\cal M};{\mathfrak D}_c\ra=|\Psi;{\cal M}\ra\times |{\mathfrak D}_c\ra$ of a {\it single} coherence domain ${{\mathfrak D}_c}$ associated with an extended object as the memory system is the direct product of the Fock space of photons and quasi-particles ${\mathfrak H}_{F,{\rm qft}}^{({{\mathfrak D}_c})}$ and the quantum mechanical Hilbert space ${\mathfrak H}_{Q,{\rm qm}}^{({{\mathfrak D}_c})}$:\footnote{For accounts of Eq. (\ref{eq:FQ}), see Sect. 1 of Ref. \cite{JMP3}.}
\begin{eqnarray}
{\mathfrak H}_{CD}^{({{\mathfrak D}_c})}={\mathfrak H}_{F,{\rm qft}}^{({{\mathfrak D}_c})}\times {\mathfrak H}_{Q,{\rm qm}}^{({{\mathfrak D}_c})}\;.\label{eq:FQ}
\end{eqnarray}
The ket vector $|\Psi;{\cal M}\ra$ belongs to ${\mathfrak H}_{F,{\rm qft}}^{({{\mathfrak D}_c})}$, and the ket vector $|{\mathfrak D}_c\ra$ belongs to ${\mathfrak H}_{Q,{\rm qm}}^{({{\mathfrak D}_c})}$.

\item[Step 2] Second, in an {\it entangled} part of the QED world $W_{\rm ent}$, there is a non-selective measurement process of an {\it entangled} superposition of memory patterns created by boundary conditions, which are {\it superposed} and are {\it entangled}, on extended objects and the electromagnetic field (refer to Fig. 4):
\begin{eqnarray}
\widehat{\varrho}_{W_{\rm ent}}&=&
\Biggl(\sum_n c_n|\{{\cal B}\}_n\ra|{0};\{{\cal M}\}_n;\{{\mathfrak D}_c\}_n\ra\Biggr)\Biggl(\sum_{n^\prime}\bar{c}_{n^\prime}\la {0};\{{\cal M}\}_{n^\prime};\{{\mathfrak D}_c\}_{n^\prime}|\la \{{\cal B}\}_{n^\prime}|\Biggr)\nonumber\\
&&\longrightarrow\sum_n |c_n|^2|\{{\cal B}\}_n\ra|{0};\{{\cal M}\}_n;\{{\mathfrak D}_c\}_n\ra\la {0};\{{\cal M}\}_n;\{{\mathfrak D}_c\}_n|\la \{{\cal B}\}_n|\label{eq:non}
\end{eqnarray}
for the quantum mechanical states $\{|\{{\mathfrak D}_c\}_n\ra\}$ of the extended objects in $\bigotimes_{{{\mathfrak D}}_c}{\mathfrak H}_{Q,{\rm qm}}^{({\mathfrak D}_c)}$.
(Here, we denote a simultaneous eigenstate, such as $|X_n^1,X_n^2,\ldots,X_n^l\ra$, by $|\{X\}_n\ra$.)

This process, namely, the formation of a mixture, automatically follows due to the unitary inequivalence of the coherent subspaces of ${\mathfrak H}_{F,{\rm qft}}^{({{\mathfrak D}_c})}$ classified by ${\cal M}$---exactly as occurs for the superselection sectors in a superselection rule for the classical field ${\cal M}({x},t)$---due to the spontaneous symmetry breaking in {\it quantum field theory}, which treats an infinite number of degrees of freedom.

\item[Step 3] Third, the {\it retrieval} of a generic written memory pattern $\{{\cal M}\}_n$ occurs by the excitation of a gapless Goldstone mode of the broken global SRS from the ground state $|0;\{{\cal M}\}_n\ra$ by perturbations of the narrow-sense boundary condition on the coherence domains as external stimuli into this system.
This gaplessness is assured by the Nambu-Goldstone theorem \cite{Umezawa}.

This process can be expressed as the change of the density matrix $\widehat{\varrho}_{W_{\rm ent}}$
\begin{eqnarray}
\widehat{\varrho}_{W_{\rm ent}}&=&
\sum_n|c_n|^2|\{{\cal B}\}_n\ra|0;\{{\cal M}\}_n;\{{\mathfrak D}_c\}_n\ra\la 0;\{{\cal M}\}_n; \{{\mathfrak D}_c\}_n|\la \{{\cal B}\}_n|\nonumber\\
&&\longrightarrow \sum_n |c_n|^2|\{{\cal B}\}_n\ra|{\Psi}_n;\{{\cal M}\}_n;\{{\mathfrak D}_c\}_n\ra \la {\Psi}_n;\{{\cal M}\}_{n}; \{{\mathfrak D}_c\}_{n}|\la \{{\cal B}\}_n|\label{eq:STEP2}
\end{eqnarray}
for the extended object states $\{|\{{\mathfrak D}_c\}_n\ra\}$ and the Goldstone boson (polariton) states $\{|{\Psi}_n;\{{\cal M}\}_n\ra\}$.
In Eq. (\ref{eq:STEP2}), {\it superposed} external stimuli (traced out as an environment) supply energy to the {\it open} quantum system of coherence domains.
Here, we suppose that each state $|{\Psi}_n;\{{\cal M}\}_n\ra$ is distinct from the others with respect to energy.

\item[Step 4] Fourth, the results of these {\it objective} processes are {\it subjectively} {\it read} by the mutually entangled coherence domains.
This process of {\it event reading} is
\begin{eqnarray}
\widehat{\varrho}_{W_{\rm ent}}&=&
\sum_n |c_n|^2|\{{\cal B}\}_n\ra|{\Psi}_n;\{{\cal M}\}_n;\{{\mathfrak D}_c\}_n\ra\la {\Psi}_n;\{{\cal M}\}_n; \{{\mathfrak D}_c\}_n|\la \{{\cal B}\}_n|\nonumber\\
&&\longrightarrow |\{{\cal B}\}_{n_0}\ra|{\Psi}_{n_0};\{{\cal M}\}_{n_0};\{{\mathfrak D}_c\}_{n_0}\ra\la {\Psi}_{n_0};\{{\cal M}\}_{n_0}; \{{\mathfrak D}_c\}_{n_0}|\la\{{\cal B}\}_{n_0}|\;,\label{eq:sel}
\end{eqnarray}
when we suppose the state-reduction mechanism {\it in the coherence domain} \cite{arXiv2}.

\end{enumerate}

This measurement scheme is a {\it type II selective measurement} in the classification of selective measurements proposed in Ref. \cite{JSTAT}.
Namely, the measuring system $\{{\cal M}\}$ that reads measurement events is inseparable from the measured system $\{{\cal B}\}+\{{\cal M}\}$ on which the non-selective measurement process (step 2) acts.

Therefore, due to the result in Ref. \cite{JSTAT}, the event reading in the fourth process requires no internal work (i.e., no energy transfer): the event reading in this scheme is an {\it unphysical} process.

With respect to step 3, we make two remarks according to Ref. \cite{Yasue}.

First, in the energy spin system coupled to the electric dipole moment ${S}^a_{\rm b.c.}$ of external stimuli as perturbations of the narrow-sense boundary condition, if the electric dipole moment ${S}^a_{\rm b.c.}$ or the strength of its coupling to electric dipole moments of elements of the system is small, or if the time scale for creating the narrow-sense boundary condition is short (namely, the perturbation energy is small), then this stimulus excites the gapless Goldstone mode and the gapped mode without changing the ground state $|0(A_c,s_c)\ra$.

In contrast, if the perturbation energy is large enough, then the non-perturbative ground state $|0(A_c,s_c)\ra$ is rewritten to another state $|0(A_c^\prime,s^\prime_c)\ra$: {\it the memory $\{{\cal M}\}$ is rewritten as a new memory $\{{\cal M}^\prime\}$.}

\section{Summary and a thought experiment}

\subsection{Summary}

In this paper, we have investigated two resonant QED models for which a quantum measurement scheme exists.

\begin{enumerate}
\item[M1] The first model is quantum measurement by the human brain involving sensory organs, which is reduced to a resonant QED system of {\it photons} and {\it ion-solvated water} \cite{PLA1,PLA2,arXiv1}.

This model is compatible with information processing in a neuron-synapse non-linear network in which the resting/firing state of each neuron is expressed by one classical bit.

\item[M2]
The second model is quantum measurement by an assembly of {\it coherence domains} with a memory function that consists of three physical processes: {\it writing}, {\it retrieval} and {\it reading}.

In the second model, every assembly of coherence domains belongs to an {\it entangled} part of the QED world.

\end{enumerate}

In both models, the external world of the measuring system is measured by the measuring system (i.e., the system of coherence domains) and the quantum measurement scheme consists of three steps: non-selective measurement, reflecting stimuli to the measuring system and event reading with information generation with respect to the event that has occurred.

These two models have a difference that results in event reading being a {\it physical} process in the first model and an {\it unphysical} process in the second model.

However, these two models are analogous to each other, with the following three correspondences:

\begin{enumerate}
\item[C1] The first correspondence is between the quantum state $|\{{\cal F}\}\ra$ of the neural network with a neural firing pattern $\{{\cal F}\}$ and the excited quantum state $|\Psi;\{{\cal M}\}\ra$ of the assembly of coherence domains $\{{{\mathfrak D}_c}\}$ with a memory pattern $\{{\cal M}\}$ (more precisely, a narrow-sense memory pattern $\{{\mathscr M}\}$ defined in ``Appendix B'') as the states of the {\it measuring systems}.

These states are the internal states in quantum measurement.

\item[C2] The second correspondence is between the sensory organ in the brain and the non-perturbative QED ground state with a spontaneous symmetry breaking.

Both of these induce non-selective measurements in their respective models and translate the stimuli into the states $|\{\cal F\}\ra$ for the first model and $|\Psi;\{{\cal M}\}\ra$ for the second model.

\item[C3] The third correspondence is between the stimulus states $|{\cal O}\ra$ and the boundary conditions on coherence domains $|\{{\cal B}\}\ra$.

These are the sources of information to be transduced, in other words, the {\it measured objects} in the two models.

\end{enumerate}

Besides these correspondences, most significantly, both measuring systems consist of {\it coherence domains}.
Due to this fact, the framework of Ref. \cite{arXiv2} used to derive the state-reduction mechanism (i.e., the event reading process in the quantum measurement scheme) can be applied to these models.

\subsection{A thought experiment}

So far, we have studied only information transduction processes.
Now, we compare the information processing in these two models.

In the first model, information processing, that is, changes in the neural states expressed using classical bits is done within the measuring system after transduction of the external stimuli into the neural firing patterns.

In contrast, in the second model, information processing is done within the external quantum mechanical world before information transduction.

Finally, based on this comparison of our models, we consider a thought experiment.

We define the {\it outage state} of the first-type measuring system $M_{\rm 1st}$ to be the situation in which {\it all neural states are the resting state $|{\cal R}\ra$ identically and sensory transduction no longer works}.

The schematic view of informational activity in the first-type measuring system $M_{\rm 1st}$ before its outage is
\begin{equation}
\xymatrix{{{{W}}}_{\rm qm}\ar@[black]@(ul,dl)[]|{{\rm IP}^1}\ar@[black][rr]^{{\rm IT}^2}&&{{M_{{\rm 1st}}}}\ar@[black]@(ru,rd)|{{\rm IP}^3}}\;,\label{eq:1st}
\end{equation}
where $W_{\rm qm}$ refers to the external quantum mechanical world, IP refers to information processing in a discrete dynamical system (that is, $W_{\rm qm}$ or $M_{\rm 1st}$) with its own time-evolution rule, and IT refers to information transduction including information reduction.

After the outage of the system $M_{{\rm 1st}}$, three facts about $M_{\rm 1st}$ follow:
\begin{enumerate}
\item[F1] $M_{{\rm 1st}}$ no longer measures any external stimulus.
Namely, $M_{{\rm 1st}}$ no longer undergoes the processes (\ref{eq:CSSR}) to (\ref{eq:dec2}) at all.
\item[F2] The neural states (i.e., $|\{{\cal R}\}_{\rm all}\ra$) have an information capacity of zero bits.
\item[F3] The synaptic strength loses its role as memory.
\end{enumerate}
Given these three facts, $M_{\rm 1st}$ has completely lost its architectural internal structure (\ref{eq:1st}) against the quantum mechanical world $W_{\rm qm}$, and $M_{\rm 1st}$ is being a part of the quantum state of the system $W_{\rm qm}$.
Then, three consequences follow:
\begin{enumerate}
\item[C1] Such a quantum state is evolved by the time-evolution rule in the discrete dynamical system $W_{\rm qm}$.
\item[C2] Such a quantum state is automatically coded by the two-fold threshold structure (\ref{eq:Condition1}) (for the density profile) and ({\ref{eq:Condition2}}) (for the temperature profile) and is stocked as a memory in a binarily reduced form (see ``Appendix B'').
This memory is stable because ${\cal E}<{\cal E}_0$ holds.
\item[C3] When we suppose the state-reduction mechanism in the coherence domain \cite{arXiv2}, this stable memory is {\it subjectively} read by the second-type measuring system $M_{\rm 2nd}$.
\end{enumerate}

Here, the schematic view of informational activity in the second-type measuring system $M_{\rm 2nd}$ is
\begin{equation}
\xymatrix{{{{W}}}_{\rm qm}\ar@[black]@(ul,dl)[]|{{\rm IP}^1}\ar@[black][rr]^{{\rm IT^1\ (coding)}}&&{{M_{{\rm 2nd}}}}}\;,\label{eq:M2nd}
\end{equation}
where the automatically coded state of $W_{\rm qm}$, with reduced generated information, is {\it directly} read (i.e., without being processed) by $M_{\rm 2nd}$.

A scheme for quantum measurement that is essentially the same as Eq. (\ref{eq:M2nd}) can be realized in a first-type measuring system $M_{\rm 1st}$ {\it before its outage} as
\begin{equation}
\xymatrix{M_{\rm 1st}\ar@[black]@(ru,rd)|{{\rm IP}}}\;,
\end{equation}
where the information transduction of $W_{\rm qm}$ into $M_{\rm 1st}$ shown in Eq. (\ref{eq:1st}) is fully blocked; then event reading is the final step in a type II selective measurement and is an {\it unphysical} process (i.e., no energy transfer accompanies event reading) \cite{JSTAT}.

We note that, for information processing in the quantum mechanical world $W_{\rm qm}$ viewed as a {\it discrete dynamical system}, there is conic space-time locality with respect to information propagation (in the sense of quantum correlation propagation), including quantum entanglement propagation for the quantum measurement process in $M_{\rm 2nd}$, despite our non-relativistic treatment of $W_{\rm qm}$.

Actually, in the spatial lattice system of $W_{\rm qm}$ in which the Hilbert space at each site is finite-dimensional and the interaction is short-range and spatially translationally invariant, it has been mathematically proved \cite{LR} and experimentally demonstrated \cite{Nature3} that there is an upper bound for the information propagation speed.

Regarding the information processing in $W_{\rm qm}$, this speed of causal interactions among elements of $W_{\rm qm}$ is analogous to the propagation speed of action potentials in information processing in a first-type measuring system $M_{\rm 1st}$.

\begin{appendix}
\section{Derivation of Eq. (37) from Eq. (36)}

In this Appendix, we derive the decoherence mechanism in Eq. (\ref{eq:CSSR}), which supports the non-selective measurement part of the measurement scheme in the first model, from Eq. (\ref{eq:gg1}).
In the equations, we denote the pair of indices $(i,j)$ by $I$.
The following calculations accord with the ideas in Ref. \cite{Araki1}.

We denote the initial quantum state vector in the Hilbert space of the combined system ${\mathfrak H}_{st}\otimes {\mathfrak H}_{so}$ at $t=t^{(1)}_{\rm ini}$ by
\begin{eqnarray}
|\Psi_0\ra&=&|\psi\ra|\varphi\ra\;,\label{eq:Psi}\\
|\psi\ra&=&\sum_n c_n|{\cal O}_n\ra\;,\\
|\varphi\ra&=&\bigotimes_I|\varphi_I\ra\;.
\end{eqnarray}
Here, $|\psi\ra$ and $|\varphi\ra$ belong to ${\mathfrak H}_{st}$ and ${\mathfrak H}_{so}$, respectively, and $\la \varphi_I|\varphi_I\ra=1$ holds for all $I$.

Then, due to the direct integral structure of the Hilbert space ${\mathfrak H}_{st}\otimes {\mathfrak H}_{so}$
\begin{eqnarray}
{\mathfrak H}_{st}\otimes {\mathfrak H}_{so}=\int^\bigoplus {\mathfrak H}_{st}(\overline{P_I})\prod_Id\overline{P_I}\label{eq:Deg2}
\end{eqnarray}
by the continuous superselection rule observables $\Bigl\{\overline{\widehat{P}_I}\Bigr\}$, the density matrix of the state vector (\ref{eq:Psi}) at $t=t_{\rm ini}^{(1)}$ is
\begin{eqnarray}
|\Psi_0\ra\la \Psi_0|&=&\int^\bigoplus |\Psi_0(\overline{P_I})\ra\la \Psi_0(\overline{P_I})|\prod_Id\overline{P_I}\label{eq:element0}\\
&=&\sum_{m,n}c_m\bar{c}_n\int^\bigoplus |{\cal O}_m\ra\la {\cal O}_n|\Biggl\{\prod_I|\varphi_I(\overline{P_I})|^2d\overline{P_I}\Biggr\}\label{eq:element}
\end{eqnarray}
and this density matrix at $t=t^{(\ell)}_{\rm fin}$ is
\begin{eqnarray}
|\Psi_{t^{(\ell)}_{\rm fin}}\ra\la \Psi_{t^{(\ell)}_{\rm fin}}|&=&\int^\bigoplus |\Psi_{t^{(\ell)}_{\rm fin}}(\overline{P_I})\ra\la \Psi_{t^{(\ell)}_{\rm fin}}(\overline{P_I})|\prod_Id\overline{P_I}\label{eq:element1}\\
&=&\sum_{m,n}c_m\bar{c}_n\int^{\bigoplus}e^{-\frac{{\rm i}}{\hbar}\sum_I\Lambda^{(i)} ({\cal E}_j({\cal O}_m)-{\cal E}_j({\cal O}_n))\overline{P_I}\delta t^{(i)}}|{\cal O}_m\ra\la {\cal O}_n|
\Biggl\{\prod_I|\varphi_I(\overline{P_I})|^2d\overline{P_I}\Biggr\}\;.
\end{eqnarray}
(See footnote 4.)
Here, $\varphi_I(\overline{P_I})$ is the wave function of the state vector $|\varphi_I\ra$ in the $\overline{P_I}$-representation.

Now, for an arbitrarily given observable of the system ${\mathfrak H}_{st}\otimes {\mathfrak H}_{so}$, selected by the continuous superselection rule,
\begin{eqnarray}
\widehat{{\cal X}}=\int^{\bigoplus}\widehat{{\cal X}}(\overline{P_I})\prod_Id\overline{P_I}\;,\label{eq:A8}
\end{eqnarray}
where $\widehat{{\cal X}}(\overline{P_I})$ acts in ${\mathfrak H}_{st}(\overline{P_I})$, its expectation value for this density matrix is calculated by
\begin{eqnarray}
\la \widehat{{\cal X}}\ra&=&\int \la \Psi_{t^{(\ell)}_{\rm fin}}(\overline{P_I})|\widehat{{\cal X}}(\overline{P_I})|\Psi_{t^{(\ell)}_{\rm fin}}(\overline{P_I})\ra \prod_I d\overline{P_I}\\
&=&\sum_{m,n}c_m\bar{c}_n\int e^{-\frac{{\rm i}}{\hbar}\sum_I\Lambda^{(i)} ({\cal E}_j({\cal O}_m)-{\cal E}_j({\cal O}_n))\overline{P_I}\delta t^{(i)}}
\Biggl\{\prod_I\la {\cal O}_n|\widehat{{\cal X}}(\overline{P_I})|{\cal O}_m\ra |\varphi_I(\overline{P_I})|^2d\overline{P_I}\Biggr\}\;.\label{eq:damp}
\end{eqnarray}
We assume that the $\overline{P_I}$-uncertainty of $|\varphi_I(\overline{P_I})|^2$ is $\Delta P_0^{(i)}$.
When the criterion (\ref{eq:gg1}) is satisfied, the damping factor in Eq. (\ref{eq:damp}) is evaluated by
\begin{eqnarray}
\Biggl|\int e^{-\frac{{\rm i}}{\hbar}\sum_I\Lambda^{(i)} ({\cal E}_j({\cal O}_m)-{\cal E}_j({\cal O}_n))\overline{P_I}\delta t^{(i)}}\Biggl\{\prod_I|\varphi_I(\overline{P_I})|^2d\overline{P_I}\Biggr\}\Biggr|
=\prod_I\frac{\Bigl|\sin(\delta^2\overline{ Q_j^{(i)}}\Delta P_0^{(i)}/\hbar)\Bigr|}{\Bigl|\delta^2\overline{ Q_j^{(i)}}\Delta P_0^{(i)}/\hbar\Bigr|}\ll 1\;,
\end{eqnarray}
where the product is taken over the sensory cells which transduce the stimulus ${\cal O}$, and $\delta^2\overline{Q_j^{(i)}}=\Lambda^{(i)}({\cal E}_j({\cal O}_m)-{\cal E}_j({\cal O}_n))\delta t^{(i)}$ is used.
So, we obtain
\begin{equation}
\la \widehat{{\cal X}}\ra \sim\sum_n|c_n|^2\int \Biggl\{\prod_I\la {\cal O}_n|\widehat{{\cal X}}(\overline{P_I})|{\cal O}_n\ra |\varphi_I(\overline{P_I})|^2d\overline{P_I}\Biggr\}\;.\label{eq:Result}
\end{equation}

The result (\ref{eq:Result}) for all observables $\widehat{{\cal X}}$ means that two state vectors belonging to different eigenspaces of $\widehat{{\cal O}}$ have no observable quantum interference.
Namely, the density matrix is equivalent to
\begin{equation}
\sum_n|c_n|^2\int^{\bigoplus}|{\cal O}_n\ra\la {\cal O}_n|\Biggl\{\prod_I|\varphi_I(\overline{P_I})|^2d\overline{P_I}\Biggr\}\label{eq:AppResult}
\end{equation}
at $t=t^{(\ell)}_{\rm fin}$.

This result leads to the non-selective measurement (\ref{eq:CSSR}).

\section{Informational interpretation of the second model}

In this Appendix, we give a brief account of an informational interpretation of the second model.

In this interpretation, the term {\it memory} is used in a narrow sense: {\it the stocking of a state, with generated information, to be read}.
To simplify the argument, we consider an idealized situation in which we can divide a given spatial domain $V$ into elementary optical domains each having the dimensions of the resonant optical wavelength $l_c$.
There, {\it states} are expressed by using classical bits where {\it the coherence domain with a superradiant phase transition/normal domain is expressed by one classical bit for each elementary optical domain} (refer to Fig. 2).

\begin{figure}[htbp]
\begin{center}
\includegraphics[width=0.6 \hsize,bb=0 0 260 109]{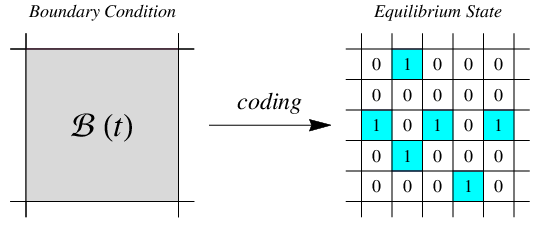}
\end{center}
\caption{The time-dependent boundary condition ${\cal B}(t)$, including the local thermal conditions on a spatial domain $V$, is coded by using $N$-bits where $N\approx V/l_c^3$ is the number of elementary optical domains in $V$.
In the coding equilibrium state, each cube with a $0$ represents a normal domain (not satisfying Eqs. (\ref{eq:Condition1}) and (\ref{eq:Condition2})) and each cube with a $1$ represents a coherence domain (satisfying Eqs. (\ref{eq:Condition1}) and (\ref{eq:Condition2})).
Normally, $0$ states form the overwhelming majority, and $1$ states are an exceptional minority.
In this sense, the repertoire of possible states and the information are actually greatly reduced.}
\end{figure}

In this language, the {\it writing} of a narrow-sense memory pattern $\{{\mathscr M}\}$ (where ${\mathscr M}$ refers to $0$ or $1$) in the (thermal) equilibrium state is the {\it process of coding} the time-dependent boundary condition ${\cal B}(t)$ including the local thermal conditions on $V$ by using $N$-bits, where $N\approx V/l_c^3$ is the number of elementary optical domains in $V$ (see Fig. 5).
Of course, the narrow-sense memory pattern $\{{\mathscr M}\}$ can be {\it rewritten} as a distinct narrow-sense memory pattern $\{{\mathscr M}^\prime\}$ in the equilibrium state by a distinct boundary condition at another time.

In quantum measurement, these two processes, that is, the writing and rewriting of a narrow-sense memory pattern, are done for a superposed component that corresponds to each superposed component of the given measured quantum state of the boundary condition, separately.

The informational content of quantum measurement lies in these two processes.
The following three processes, that is, the non-selective measurement, the retrieval of memory, and the reading of memory are the physical procedures in quantum measurement.

\begin{enumerate}
\item[P1] First, the non-selective measurement process in the second model is attributed to spontaneous symmetry breaking occurring in the coherence domains and is for the entangled parts of the ground state.

\item[P2]
Second, the {\it retrieval} of the memory pattern is done by the injection of energy, as an external stimulus, into the system $V$ (i.e., the excitation of the ground state of $V$).

Here, note that although the gaplessness in the excitation energy spectrum from the ground state energy is not ensured for the normal domain, the excitation energy spectrum in the coherence domain is gapless due to the Goldstone theorem.

After this memory retrieval process, the entangled part of the memory in the excited state is ready to be read.

\item[P3]

Finally, in our supposition of the results in Ref. \cite{arXiv2}, the reading of memory is attributed to the coherence domains and done for the mixture of states excited from the ground state of $V$.

\end{enumerate}

In summary, the second-type measuring system has the ability to transduce information generated by the external quantum mechanical world and perform subsequent quantum measurement but does {\it not} involve information processing within the system.

\end{appendix}

\end{document}